\documentclass[aps,10pt,prd,twocolumn,superscriptaddress,preprintnumbers,%
showkeys,nofootinbib,showpacs]{revtex4-2}
\usepackage{amsmath,amsfonts,amssymb,amscd,amsxtra,amsthm}
\usepackage{graphicx}  
\usepackage{epstopdf}
\usepackage{dcolumn}  
\usepackage{bm}          
\usepackage{slashed}
\usepackage{cancel}
\usepackage{float}
\usepackage{mathtools}
\usepackage{amsbsy}
\usepackage{amstext}
\usepackage{hyperref}

\usepackage{wasysym}
\usepackage[utf8]{inputenc} 
\usepackage{booktabs} 
\usepackage[normalem]{ulem} 
\usepackage[dvipsnames]{xcolor} 
\usepackage{tabularx}
\usepackage{enumitem}  
\usepackage{array} 
\usepackage{slashed}
\usepackage{tikz}
\usepackage{float}
\usepackage{multirow}
\usepackage{cancel}
\usepackage{physics}
\renewcommand\sout{\bgroup \color{red} \ULdepth=-.5ex \ULset}

\makeatletter

\begin{document}  
\preprint{INHA-NTG-02/2025}
\title{Nucleon tensor form factors at large $N_{c}$}  

\author{Nam-Yong Ghim}
\email[E-mail: ]{Namyong.ghim@inha.edu}
\affiliation{Department of Physics, Inha University, Incheon 22212,
  Republic of Korea} 

\author{Ho-Yeon Won}
\email[E-mail: ]{hoyeon.won@polytechnique.edu}
\affiliation{CPHT, CNRS, \'Ecole polytechnique, Institut Polytechnique
  de Paris, 91120 Palaiseau, France} 

\author{June-Young Kim}
\email[E-mail: ]{jykim@jlab.org}
\affiliation{Theory Center, Jefferson Lab, Newport News, VA 23606,
  USA} 

\author{Hyun-Chul Kim}
\email[ E-mail: ]{hchkim@inha.ac.kr}
\affiliation{Department of Physics, Inha University, Incheon 22212,
  Republic of Korea} 
\date{\today}
\begin{abstract}
We investigate nucleon tensor form factors in the large-$N_{c}$
limit. In this picture, the nucleon emerges as a state of the
$N_c$ valence quarks, which were bound by pion mean fields that were
created by the presence of the valence quarks self-consistently. 
We find that the tensor charge ($g^{u-d}_{T}=0.99$)
and the anomalous tensor magnetic moment ($\kappa^{u+d}_{T}=7.61$) are
dominated by valence quarks, while the tensor quadrupole moment 
($Q^{u-d}_{T}=-7.02$) shows significant sea quark effects. We examine
how these quantities vary as the average size of the pion mean field is
changed, showing interpolation between non-relativistic quark and
Skyrme limits. We also observe that $g^{u-d}_{T}$
and $\kappa^{u+d}_{T}$ depend weakly on the pion mass. In contrast,
$Q^{u-d}_{T}$ exhibits strong enhancement near the chiral limit. The
numerical results are in good agreement with available lattice QCD
data and provide predictions for unmeasured quantities.   
\end{abstract}
\maketitle
\tableofcontents

\section{Introduction}
Understanding the nucleon structure has been a fundamental issue in 
hadronic physics well over decades. Generalized parton
distributions~(GPDs) have emerged as a crucial tool in examining
nucleon structure within quantum chromodynamics~(QCD)~\cite{Ji:1998pc,
  Goeke:2001tz, Diehl:2003ny, Belitsky:2005qn, Boffi:2007yc}. GPDs
parametrize nucleon matrix elements of quark and gluon light-ray
operators at non-zero momentum transfer, unifying the concepts of
parton distributions functions (PDFs) and form factors. Chiral-even
structures have extensively studied and well understood. The
chiral-even GPDs can be extracted experimentally from deeply virtual
Compton scattering and hard-exclusive vector meson
production~\cite{Ji:1998pc, Goeke:2001tz, Diehl:2003ny,
  Belitsky:2005qn, Boffi:2007yc, Guidal:2013rya, Kumericki:2016ehc,
  dHose:2016mda}.  
In contrast, chiral-odd structures cannot be measured through these
processes, and alternative methods proposed~\cite{Collins:1999un,
  Ivanov:2002jj, Enberg:2006he, Pire:2009ap, ElBeiyad:2010pji,
  Pire:2015iza, Pire:2017lfj, Duplancic:2023kwe, Ahmad:2008hp,
  Goloskokov:2009ia, Goloskokov:2011rd, Goloskokov:2013mba} face
significant challenges. Consequently, dynamical information on
chiral-odd GPDs, primarily based on model-dependent
studies~\cite{Pasquini:2005dk, 
  Burkardt:2007xm, Lorce:2007fa, Chakrabarti:2008mw, Kumar:2015yta,
  Chakrabarti:2015ama, Tezgin:2024tfh}, remains limited. 
Despite these experimental and theoretical challenges, understanding
chiral-odd GPDs is as equally important as their chiral-even
counterparts for a comprehensive mechanical interpretation of nucleon
structure. In the leading twist, four chiral-odd GPDs quantify the 
distribution of transversely polarized quarks within the
nucleon~\cite{Diehl:2005jf}. 
Form factors for the antisymmetry tensor operator can be interpreted as the
first Mellin moments of chiral-odd GPDs and will provide a constraint
on the chiral-odd GPDs/PDFs. These moments are potentially extractable
from polarization observables in semi-inclusive deep-inelastic
scattering and, in principle, from dilepton production in polarized
proton-proton collisions~\cite{Ralston:1979ys, Jaffe:1991ra,
  Jaffe:1991kp, Collins:1992xw}. However, these methods may result in 
substantial theoretical and experimental
uncertainties~\cite{PAX:2005leu, Anselmino:2007fs, Anselmino:2008jk,
  Bacchetta:2012ty, Anselmino:2013vqa, Kang:2014zza}. Lattice QCD
simulations have provided results for the lowest $x$-moment of
chiral-odd GPDs~\cite{QCDSF:2006tkx, Park:2021ypf,
  Alexandrou:2021oih}. Beyond these computational estimates and the
aforementioned experimental approaches, current knowledge of tensor
form factor properties remains limited. 

Spontaneous breakdown of chiral symmetry (SB$\chi$S) in QCD can be
realized by the topological fluctuations of gauge fields in the
instanton vacuum; see Refs.~\cite{Schafer:1996wv, Diakonov:2002fq} for
a review. Instantons, solutions to the Yang-Mills equation in singular
gauge and Euclidean time, exhibit characteristic dimensions: average
size $\bar{\rho}\approx 1/3$ fm and mean separation $\bar{R}\approx 1$ 
fm. Instantons generate localized quark zero modes with definite
chirality. These modes are subsequently delocalized and undergo
chirality flips when quarks propagate and interact consecutively with
instantons and antiinstantons through the instanton
medium. This process causes SB$\chi$S. This picture
introduces a small parameter called the instanton packing fraction
that enables a systematic expansion for studying SB$\chi$S and hadronic
correlation functions. Effective dynamics at the scale $\bar{R}$ can
be constructed using $1/N_c$ expansion techniques, including saddle
point approximation and bosonization. This approach yields a
description of quarks with the dynamical mass $M \sim 0.3-0.4$ GeV
coupled to a chiral pion field. Within this framework, the nucleon
emerges as a self-consistent mean-field solution characterized by a
classical pion field (soliton) and quarks occupying single-particle  
states~\cite{Diakonov:1987ty}. This mean-field approach provides a
specific realization of the general baryon picture in large-$N_c$
QCD~\cite{Witten:1979kh}. It represents an interpolation between quark
model and chiral soliton~\cite{Zahed:1986qz} descriptions,
encompassing both as limiting cases~\cite{Praszalowicz:1995vi,
  Praszalowicz:1998jm}. The picture of an effective chiral theory for
the nucleon has been successfully applied to desribing various
hadronic observables; see Ref.~\cite{Christov:1995vm} for a 
review. 

In this paper, we employ an effective chiral theory in the large $N_c$
limit to investigate the nucleon matrix element of the local
antisymmetric tensor current,  based on a previously 
described framework that preserves important QCD properties, such as
sum rules and polynomial properties, for chiral-odd
GPDs~\cite{Kim:2024ibz}. In addition, this approach~\cite{Kim:2024ibz}
has provided valuable insights into the nucleon's spin-flavor
structure for the matrix element of the chiral-odd partonic operator.  
In this work, we present numerical estimations of tensor form factors,
systematically expanding in terms of the multipole order. Unlike previous
studies~\cite{Kim:1995bq, Kim:1996vk, Ledwig:2010zq}, the current
analysis is confined to the strict large $N_c$ limit of QCD, ensuring
consistency between chiral-odd GPDs~\cite{Kim:2024ibz} and tensor form
factors. We will first revisit the tensor monopole ($g_T$) and
tensor dipole ($\kappa_T$) form factors, offering more transparent
insights into their dynamical information and correcting previous
calculations of the tensor dipole form factors. We also 
explore the previously uninvestigated quadrupole structure of the
tensor form factors ($\sim Q_T$). In addition to the tensor charge
$g_T$ and tensor anomalous magnetic moment $\kappa_T$, the quadrupole
moment $Q_T$ is equally important, which is also related to the
leading-twist chiral-odd GPDs $\tilde{H}_{T}$.  Then we examine the
large-$r$ behavior of the  three-dimensional~(3D) tensor distributions
and investigates the pion mass dependence to elucidate the chiral
properties of the tensor form factors. This comprehensive analysis 
provides a more complete and accurate description of the nucleon
tensor structure within the framework of the effective chiral
theory.

The paper is organized as follows: Section II introduces the formalism
for tensor form factors and we perform the mutipole expansion in
specific frames. Section III details the effective chiral theory and
our calculation methods. Sections IV and V present our numerical
results and discussion, respectively. Finally, we summarize our
findings and offer concluding remarks in Section VI. 

\section{Tensor form factors}
The nucleon matrix element of the local antisymmetric tensor operator
is given by 
\begin{align}
\mathcal{M}^{q}[i\sigma^{\mu \nu}]&=  \langle N (p',s') |
   \bar{\psi}_{q}(0) i\sigma^{\mu \nu} \psi_{q} (0) | N(p,s) 
   \rangle,  
\label{eq:General_ME_tensor0}
\end{align}
where $\psi_{q}$ denotes the quark field operator with its flavor
component $q=u,d, \ldots$ and $\sigma^{\mu \nu} = \frac{i}{2}
[\gamma^{\mu}, \gamma^{\nu}]$. The nucleon state is normalized to be
$\langle  N(p',s') | N(p,s) \rangle = (2\pi)^{3} 2p^{0}
\delta_{s's} \delta^{(3)} (\bm{p}'-\bm{p})$, where $p'(p)$
denotes the final~(initial) state momentum, and $s'(s)$ represents
the spin projection of the final~(initial) nucleon state.
Using the Lorentz covariance and discrete symmetries (time reversal,
parity, hermiticity), the matrix element~\eqref{eq:General_ME_tensor0}
is parameterized in terms of the three independent tensor form factors   
\begin{align}
\mathcal{M}^{q}[i\sigma^{\mu \nu}]&= \bar{u} \bigg{[}H^{q}_{T} i
  \sigma^{\mu \nu} +  \tilde{H}^{q}_{T} \frac{P^{\mu}
  \Delta^{\nu} - \Delta^{\mu}  P^{\nu}}{M^{2}_{N}} \cr 
&\hspace{0.5cm}+ E^{q}_{T} \frac{\gamma^{\mu}\Delta^{\nu} 
-\Delta^{\mu} \gamma^{\nu} }{2M_{N}} \bigg{]} u, 
\label{eq:General_ME_tensor}
\end{align}
where $P$ and $\Delta$ denote respectively the average and difference
of the momenta 
\begin{align}
P=\frac{p'+p}{2}, \quad \Delta=p'-p.
\end{align}
Note that the Dirac gamma matrices satisfy
$\{\gamma^\mu,\,\gamma^\nu\} = 2g^{\mu\nu}$, and  
Dirac spinors are normalized to the nucleon mass $\bar{u}(p,s)
u(p,s) = 2 M_{N}$. In Eq.~\eqref{eq:General_ME_tensor}, the tensor
form factors $H^{q}_{T}\equiv H^{q}_{T}(t;\mu),
\tilde{H}^{q}_{T}\equiv \tilde{H}^{q}_{T}(t;\mu)$, and 
$\tilde{E}^{q}_{T}\equiv \tilde{E}^{q}_{T}(t;\mu)$ are given as the
real functions of the momentum transfer squared, $t=\Delta^{2}$, and
can be understood as the first Mellin moments of the chiral-odd
GPDs~\cite{Diehl:2001pm, Hagler:2004yt}. Since the tensor current is
not a conserved one, the form factors must essentially depend on the
normalization scale $\mu$. In this work we will suppress its
dependence in the expressions, e.g., $H^{q}_{T}(t):=H^{q}_{T}(t;\mu)$.  

The matrix element of the antisymmetric tensor operator can be
analyzed in two distinct reference frames: the Drell-Yan-West frame
and the 3D Breit frame. The two-dimensional~(2D) Drell-Yan-West frame
offers an unambigious spatial interpretation of the form factor. On
the other hand, the 3D Breit frame~(BF) is hampered by an ambiguous
spatial interpretation, because the wave packets of the baryon states
are not sufficiently localized. Thus, unless we treat the collective
baryon system as non-relativistic, a proper interpretation of the 3D
spatial distribution would be deviated by the relativistic
corrections$~\sim 20\%$ for the nucleon. However, in the large $N_c$
limit of QCD, where the baryon mass $M_B$ scales with $\sim N_c$, the
baryon is almost static. Thus, we can reasonably consider the 3D Breit
frame. This frame is particularly well suited for studying form
factors from the perspective of the 3D partial wave expansion
(multipole expansion). For a comprehensive understanding of tensor
form factors, we need to examine both reference frames: the 3D BF for
practical analysis in the large $N_{c}$ limit, and the 2D
Drell-Yan-West frame for clear interpretation. In the next section, we
will conduct a multipole expansion of the matrix element given in
Eq.~\eqref{eq:General_ME_tensor0} in both of these frames.  

\subsection{Three-dimensional Breit frame}
Using the standard 3D BF, where $\bm{P}=\bm{0}$,
we first study the tensor structure of the matrix elements. In this
symmetric frame, $P$ and $\Delta$ are given by 
\begin{align}
P = (P^{0}, \bm{0}), \quad \Delta = (0, \bm{\Delta}),
\label{eq:3DBF}
\end{align}
where the instant-form components are represented as $v=
(v^{0},\bm{v})$. Obviously, these momentum variables~\eqref{eq:3DBF}
satisfy the on-shell conditions 
\begin{align}
P \cdot \Delta =0, \quad P^{2} +\frac{\Delta^{2}}{4} = M^{2}_{N}.
\label{eq:onshell}
\end{align}
The angular dependence of the 3D momenta ($p, p'$) is only carried by
the momentum transfer $\bm{\Delta}$. Therefore, the 3D multipole
expansion of the matrix element~\eqref{eq:General_ME_tensor} can be
performed with respect to the momentum transfer $\bm{\Delta}$. The 3D
rank-$n$ irreducible tensors are defined by 
\begin{align}
&Y_{0} = 1 && (L=0), \nonumber \\[1ex]
&Y^{i}_{1} = \frac{\Delta^{i}}{|\bm{\Delta}|} && (L=1),\cr
&Y^{ij}_{2} = \frac{\Delta^{i}\Delta^{j}}{|\bm{\Delta}|^{2}} -
 \frac{1}{3}\delta^{ij}    && (L=2),  
\label{eq:3D_vec}
\end{align}
where $i,j$ run over the 3D components $i,j=1,2,3$. The angular
momenta $(L=0, 1, 2)$ of the irreducible tensors are indicated in
Eq.~\eqref{eq:3D_vec}. The matrix element of the nucleon depends not
only on the momentum transfer, but also it is controlled by the spin
polarizations $s'$ and $s$ of the initial and final nucleon states. 

The spin structures appear as bilinear forms in the 2-component
spinors $\chi$ describing the spin wave function of each nucleon in
its rest frame, 
\begin{align}
\chi^{\dagger}(s') \hat{M} \chi(s).
\end{align}
In the following the quantization axis is chosen as the 3-axis, and
the spinors $\chi$ are eigenspinors of the third component of the spin
operator $\bm{S}=\bm{\sigma}/{2}$, where $\bm{\sigma}$ denotes the
SU(2) Pauli matrix. Thus, the spin quantum numbers are given by the spin
projection along the 3-axis in the rest frame, 
\begin{align}
s \equiv S_{3}, \quad \quad s' \equiv S'_{3}.
\end{align}
The operator $\hat{M}$ can
be either a unit operator $\bm{1}$ or a component of the spin operator
$\bm{S}=\bm{\sigma}/{2}$. In this context, the representations of the
spin states in the 3D expansion are related to the orbital angular
momentum $L$  
\begin{align}
\hat{M}=&\bm{1} && (L=0), \cr
&\sigma^{i}  && (L=1).
\label{eq:3D_spin}
\end{align}

Using Eqs.~\eqref{eq:3D_vec} and~\eqref{eq:3D_spin}, we perform the 3D
multipole expansion of Eq.~\eqref{eq:General_ME_tensor} for the
separate 3D $0k$- and $ij$-components:  
\begin{subequations}
\label{eq:3Dmul} 
\begin{align}
\mathcal{M}^{q}[i\sigma^{0 k}] &= \bm{1} \,  Y^{i}_{1} \sqrt{-t}
  \left[ H^{q}_{T} +E^{q}_{T} + 2\tilde{H}^{q}_{T}
                    \right], \label{eq:3Dmul_a} \\ 
\mathcal{M}^{q}[i\sigma^{ij}] &=i \epsilon^{ijk} \sigma^{k} Y_{0}
  2M_{N} \left[ H^{q}_{T} +\frac{t}{6M^{2}_{N}}  E^{q}_{T}
  \right] \cr 
&-i \epsilon^{ijk} \sigma^{m} Y^{km}_{2} \frac{t}{2M_{N}} \left[
      \frac{1}{2} H^{q}_{T} +E^{q}_{T}
  \right].
\label{eq:3Dmul_b}  
\end{align}
\end{subequations}
The canonical spin states, which are constructed by boosting the spin
states at rest frame, are used in the multipole expansion of
Eqs.~\eqref{eq:General_ME_tensor0}
and~\eqref{eq:General_ME_tensor}. Here it is implied that the spin 
operators are contracted with the nucleon rest-frame spinors, and that
the matrix element is given as a function of $S_3$ and $S'_{3}$.  

There are three independent multipole structures in
Eqs.~\eqref{eq:3Dmul_a} and~\eqref{eq:3Dmul_b}. These multipole 
structures can also be explained by addition of the orbital
angular momentum. By combining Eqs.~\eqref{eq:3D_vec}
and~\eqref{eq:3D_spin}, and by considering parity and time-reversal
symmetries, we can project the operators $i\sigma^{0 k}$ and
$i\sigma^{ij}$ in such a way that they contain the multipole 
structure of the orbital angular momentum $L=1$.  
Note that this projection is consistent with Eq.~\eqref{eq:3Dmul}. For
example, the rank-2 tensor $Y_{2}$ ($L=2$) contracted with the spin
operator $\sigma^{i}$ ($L=1$) preserves the angular momentum of the
operator $i\sigma^{ij}$ ($L=1$). The parity of  this structure is
calibrated by the antisymmetric tensor $\epsilon^{ijk}$, which is
consistent with the second line of Eq.~\eqref{eq:3Dmul_b}; see
Ref.~\cite{Kim:2024ibz} for details.   

\subsection{Two-dimensional Drell-Yan-West frame}
We also examine the matrix element of the tensor operator using the
standard two-dimensional~(2D) Drell-Yan-West frame~(DYWF), satisfying
$\bm{P}_{\perp}=0$. In this symmetric frame, $P$ and $\Delta$ are
given by  
\begin{align}
P &= \left(P^{+}, \frac{4M^{2}_{N} + |\bm{\Delta}_{\perp}|^{2} }{8
    P^{+}} ,    \bm{0}_{\perp} \right), \\[1ex]
\Delta &= \left(0, 0, \bm{\Delta}_{\perp}\right),
\label{eq:2Dmo}
\end{align}
where the light-front components are represented as $v=
(v^{+},v^{-},\bm{v}_{\perp})$ and $v^{\pm} = (v^{0}\pm
v^{3})/2$. Again, these momentum variables~\eqref{eq:2Dmo} satisfy the
on-shell conditions~\eqref{eq:onshell}. The angular dependence of the
2D momenta is carried only by the momentum transfer
$\bm{\Delta}_{\perp}$. Therefore, the 2D multipole expansion of the
matrix element~\eqref{eq:General_ME_tensor} can be performed with
respect to the momentum transfer $\bm{\Delta}_{\perp}$. The 2D
irreducible rank-$n$ tensors are defined by~\cite{Kim:2022wkc} 
\begin{align}
&X_{0} = 1 &&(L_3=0), \nonumber \\[1ex]
  &X^{i}_{1} = \frac{\Delta^{i}_{\perp}}{|\bm{\Delta}_{\perp}|}
           && (L_3=\pm 1),\cr
  &X^{ij}_{2} =  \frac{\Delta_{\perp}^{i}\Delta_{\perp}^{j}}
    {|\bm{\Delta}_{\perp}|^{2}} - \frac{1}{2}\delta^{ij}
           &&(L_3=\pm 2),
\label{eq:2D_vec}
\end{align}
where $i,j$ are the transverse components $i,j=1,2$. The longitudinal
components of the orbital angular momentum $(L_{3}=0, \pm 1, \pm 2)$
for the 2D irreducible tensors are indicated in
Eq.~\eqref{eq:2D_vec}. Similar to the 3D multipole expansion, the spin
structure is represented by the 3D spin vector~\eqref{eq:3D_spin} with 
the rotational symmetry broken: 
\begin{align}
&\bm{1} && ( L_3=0 ), \cr
&\sigma^{3}  && (L_3 = 0), \cr
&\sigma^{i}  && (L_3 = \pm 1),
\label{eq:2D_spin}
\end{align}
where $i=1,2$. Using Eqs.~\eqref{eq:2D_vec} and~\eqref{eq:2D_spin}, we
perform the 2D multipole expansion of Eq.~\eqref{eq:General_ME_tensor}
for the $+j$ component: 
\begin{align}
  \mathcal{M}^{q}[i\sigma^{+j}] =2P^{+} &\bigg{[} i \epsilon^{3jm}
\sigma^{m} \, X_{0}   \left\{ H^{q}_{T} - \frac{t}{4M^{2}_{N}}
 \tilde{H}^{q}_{T}  \right\} \cr
  &+  \bm{1} \, X^{j}_{1} \frac{ \sqrt{-t}}{2M_{N}}
    \left\{ E^{q}_{T} + 2 \tilde{H}^{q}_{T} \right\}  \cr
  &+i \epsilon^{3jl} \sigma^{m} \, X^{lm}_{2}
    \frac{t }{4M^{2}_{N}} \left\{    2\tilde{H}^{q}_{T} \right\} \bigg{]}. 
\label{eq:2DLF_multipole_FF}
\end{align} 
Here, the light-front helicity states, which are constructed by
performing the light-front boost on the spin states at rest, 
are used in the multipole expansion of
Eq.~\eqref{eq:General_ME_tensor}. Similar to Eq.~\eqref{eq:3Dmul} 
the spin operators can also be contracted with the nucleon
rest-frame spinors. The matrix element of the $+j$-component of the
tensor operator consists of the 2D monopole, dipole, and quadrupole
structure, respectively, 
\begin{align}
&\left[ H^{q}_{T} - \frac{t}{4M^{2}_{N}} \tilde{H}^{q}_{T}  \right], \cr
&\left[ E^{q}_{T} + 2 \tilde{H}^{q}_{T} \right], \cr
&\left[ 2\tilde{H}^{q}_{T} \right]. 
\label{eq:multipole}
\end{align}
In the forward limit, $t\to 0$, the tensor multipole form factors
become respectively the tensor charge, the anomalous tensor magnetic
moment, and the tensor quadrupole moment: 
\begin{align}
H^{q}_{T}(0) &= g^{q}_{T}, \cr
E^{q}_{T}(0)+2\tilde{H}^{q}_{T}(0) &= \kappa^{q}_{T}, \cr
2\tilde{H}^{q}_{T}(0) &= Q^{q}_{T}.
\end{align}

The number of the multipole structure~\eqref{eq:2DLF_multipole_FF} can
also be explained by the addition of the longitudinal orbital angular
momentum. At the leading twist accuracy, the orbital angular momentum
($L_{3}=\pm1$) of the chiral-odd operator can be formed by combining 
Eqs.~\eqref{eq:2D_vec} and~\eqref{eq:2D_spin} and by considering
the discrete symmetries. The maximum rank of the orbital angular
momentum can reach the quadrupole tensor $X_{2}$ $(L_{3}=\pm 2)$
combined with the spin dipole operator $\sigma^{i}$ $(L_{3}=\pm
1)$. The parity of this multipole structure is controlled by the
antisymmetric tensor $\epsilon^{ijk}$. In this way the whole multipole
structure in Eq.~\eqref{eq:2DLF_multipole_FF} can be explained; see 
Ref.~\cite{Kim:2024ibz} for details. 

On the other hand, the orbital angular momentum of the chiral-even
operator $(\gamma^{+},\gamma^{+}\gamma_{5})$ at the leading twist is
$L_{3}=0$. Therefore, the quadrupole orbital angular momentum ($X_2$)
cannot be constructed. The possible maximum multipole is the dipole
orbital angular momentum $X_{1}$ $(L_{3}=\pm 1)$ coupled with the
dipole spin $\sigma^{i}$ $(L_{3}=\pm 1)$. This means that the
quadrupole structure appears uniquely in the nucleon matrix element of
the chiral-odd operator at the leading twist accuracy. 

One of the virtues of the multipole expansion lies in the fact that we
can easily understand the physical meaning of the form factors. The Fourier
transform of the given multipole form factors~\eqref{eq:multipole} can
be understood as the impact parameter distribution. The multipole form 
factors and the multipole pattern of the impact parameter distribution
have one-to-one correspondence. For example, the dipole
$E_{T}+2\tilde{H}_{T}$ and quadrupole $2\tilde{H}_{T}$ form factors
quantify how the transversely polarized quark distributions are
distorted in the form of the dipole and quadrupole patterns in the
impact parameter space; see Ref.~\cite{Diehl:2005jf} for
details. Therefore, the new basis~\eqref{eq:multipole} has a clear
physical interpretation and can be employed in discussion of the
nucleon structure. 

\section{Effective chiral theory}
The SB$\chi$S in QCD is realized by the topological
fluctuations of the gauge fields in the instanton vacuum. Based on
this picture, effective dynamics~\cite{Diakonov:1987ty,
  Diakonov:2002fq} at the scale $\bar{R}$ has emerged at large
$N_{c}$. This effective theory is characterized by the low-energy QCD
partition function  
\begin{align}
Z_{\mathrm{eff}} &= \int \mathcal{D} U \exp(i S_{\mathrm{eff}}), 
\label{eq:par}
\end{align}
where the effective chiral action is given by
\begin{align}
&\exp(i S_{\mathrm{eff}}[U]) =  \int \mathcal{D} \psi \mathcal{D}
                {\bar{\psi}} \cr 
&\times\exp[ \int d^{4}x \, \bar{\psi}(x) \left(i \slashed{\partial} -
                                    M U^{\gamma_{5}} - \hat{m} \right)
                                    \psi(x) ]. 
\label{eq:action}
\end{align}
$\hat{m}$ represents the current quark mass, We assume 
isospin symmetry with $m_{\mathrm{u}}=m_{\mathrm{d}}=m$, so that the
mass matrix is proportional to the unit matrix, i.e.,
$\hat{m}=\mathrm{diag}(m,m)$. This picture describes quarks with the 
dynamical mass $M$ (approximately $0.3-0.4$ GeV) coupled to a chiral
pion field $U^{\gamma_{5}}$. While $M$ generally depends on the
momentum of the quark, $\bar{\rho}^{-1} \approx 0.6$ GeV 
serves as a natural ultraviolet (UV) cutoff in this theory. For
simplicity, $M$ is treated as a constant by introducing a UV cutoff
$\Lambda\approx \bar{\rho}^{-1}$. The SU(2) chiral fields $U$ are
defined as: 
\begin{align}
&U= \exp{i \pi^{a}(x) \tau^{a}}, \cr 
&U^{\gamma_{5}} = \exp{i \pi^{a}(x) \tau^{a} \gamma_{5}}=
\frac{1+\gamma_{5}}{2} U + \frac{1-\gamma_{5}}{2} U^{\dagger}, 
\end{align}
where $\pi^{a}(x)$~$(a = 1, 2, 3)$ represents the
pseudo-Nambu-Goldstone (pNG) boson field, and $\tau^{a}$ is the SU(2)
flavor matrix. The low normalization point of 
this theoretical framework is about $600$ MeV, which is the inverse of
the average instanton size $\bar{\rho}^{-1}$ (see
Ref.~\cite{Diakonov:1995qy}).  

\subsection{Nucleon in the mean-field picture}
The nucleon in this framework emerges as a self-consistent mean-field
solution, characterized by the classical pion field and quarks in
single-particle states. This mean-field solution is obtained by using
semiclassical approximation, which is justfied at large $N_{c}$. In this
approach, developed by Diakonov et al.~\cite{Diakonov:1987ty}, the
nucleon emerges as a state consisting of the $N_c$ valence quarks
bound by the pion mean field. In this section we will give a brief
summary of this mean-field approach.  

In Euclidean space, the correlation function of baryon currents
composed of the $N_c$ valence quark fields, are computed in the
effective chiral theory~\eqref{eq:action} by employing the saddle point
approximation (semiclassical approximation) which is valid at large
$N_{c}$. The mean-field solution is characterized by a nontrivial
classical chiral field $U_{\mathrm{cl}}$, which is obtained by solving
the saddle-point equation or the classical equation of motion: 
\begin{align}
\frac{\delta S_{\mathrm{eff}}[U]}{\delta U} \bigg{|}_{U=U_{\mathrm{cl}}} = 0.
\label{eq:cEOM}
\end{align}
The solution is time-independent (static) of the Euclidean time 
and has a specific spatial configuration, where the isospin
orientation aligns with the spatial direction, forming a structure
known as a ``hedgehog''~\cite{Pauli:1942kwa}: 
\begin{align}
U_{\mathrm{cl}}(\bm{x})&= \exp{i \pi^{a}_{\mathrm{cl}}(\bm{x}) \tau^{a}},\cr
\pi^{a}_{\mathrm{cl}}(\bm{x}) &= \frac{x^{a}}{|\bm{x}|} P(r), \quad (a=1,2,3).
\label{eq:hed}
\end{align}
$P(r)$ denotes a radial profile function with $P(0) = \pi$ and
$P(\infty) = 0$, where $r \equiv |\bm{x}|$ represents the magnitude of
the position vector. This field configuration maintains invariance
under simultaneous spatial and flavor rotations, and thus embodies the
emergent spin-flavor symmetry of baryons in the large-$N_c$ limit.  

The quark field can be represented in a first-quantized
representation, where the quarks occupy the single-particle states
within the classical chiral field. These states are governed by the
one-body Dirac Hamiltonian
\begin{align}
  &H(U_{\mathrm{cl}})= -i \gamma^{0} \gamma^{k} \partial_{k}
    + \gamma^{0} M U^{\gamma_{5}}_{\mathrm{cl}} +
    \gamma^{0} \hat{m}, \nonumber \\[1ex]
&H \Phi_{n}(\bm{x})= E_{n}\Phi_{n}(\bm{x})
\label{eq:Hamil}
\end{align}
in Minkowski space. Because of the hedgehog structure mentioned above, 
the Hamiltonian commutes with the grand spin
$\bm{G}=\bm{\Sigma}/2+\bm{T}+\bm{L}$ and the parity operators 
$\hat{\Pi}$,  
\begin{align}
[H, \bm{G} ] = 0, \quad [H, \hat{\Pi} ] = 0, 
\end{align}
where $\bm{\Sigma}= -\gamma_{0}\bm{\gamma}\gamma_{5}$,
$\bm{T}=\bm{\tau}/2$, and $\bm{L}=\bm{x}\times \bm{p}$ denote the
spin, isospin, and orbital angular momentum operators,
respectively. The quark state is thus characterized by the 
corresponding quantum numbers: 
\begin{align}
&|n \rangle \equiv |n = \{E_{n}, G, G_{3}, \Pi \} \rangle, \nonumber \\[1ex]
&\text{with} \quad  \Phi_{n}(\bm{x})\equiv \langle \bm{x}| n \rangle.
\end{align}
The one-body Dirac Hamiltonian~\eqref{eq:Hamil} is solved in the
presence of the static pion field to determine the eigenfunctions
$\Phi_{n}(\bm{x})$ and eigenenergies $E_{n}$ of the massive
quark. This diagonalization is carried out within a finite-sized box,
employing the ``finite box'' method developed by Ripka and
Kahana~\cite{Kahana:1984dx, Kahana:1984be} (see also
Ref.~\cite{Christov:1995vm} for a review). The energy spectrum of the
system consists of a discrete level with energy $E_{\mathrm{lev}}<M$
and spectra of continuum levels with energies $E_{n}<M$ and
$E_{n}>M$. In the ground state of a nucleon with baryon number $B =
1$, both the discrete level and the negative continuum are occupied by
$N_c$ quarks. 

The energy of the classical nucleon, to the leading order in the
$1/N_c$ expansion, is derived by the sum of the energies of the
discrete level and the negative continuum, with the vacuum energy
subtracted:  
\begin{align}
M_{N} &= N_{c} \sum_{n, \mathrm{occ}} E_{n} - \mathrm{vac} \cr
&= N_{c} E_{\mathrm{lev}} + N_{c} \sum_{E_{n}<0} E_{n} - \mathrm{vac},
\label{eq:energy}
\end{align}
where ``occ'' encompasses both the discrete level and the negative
continuum level.  Since the resulting energy is logarithmically
divergent, it is not sensitive to a specific regularization scheme. In
this study, we use the proper-time regularization to tame the UV
divergence of the Dirac continuum (see Ref.~\cite{Christov:1995vm}). 

\subsection{Zero-mode quantization}
As mentioned previously, $1/N_c$ mesonic quantum fluctuations have
been neglected. So, the nucleon is still a classical one, which 
does not carry proper quantum numbers. The pion mean field has still
quantum fluctuations arising from translational and rotational zero
modes, which do not change the energy of the classical nucleon. Since
these zero modes are related to rotational and translational
symmetries, and are not at all small, we have to integrate over them
completely. Then they furnish the classical nucleon with proper
quantum numbers, we must integrate over the zero-mode fluctuations in
a complete manner. For the detailed explanation about the zero modes
within the framework of the pion mean-field approach, we refer to
Refs.~\cite{Diakonov:1987ty, Christov:1995vm, Diakonov:1997sj}. Here,
we recapitulate briefly the zero-mode quantization within the present
framework.   

The zero-mode quantization is summarized in terms of the following 
functional integral over the pNG fields
\begin{align}
&\int \mathcal{D}U \mathcal{F}[U(\bm{x})]  \to \cr
&\int \mathcal{D}\bm{Z} \int
  \mathcal{D} R \, \mathcal{F}\left[ T R
 U_{\mathrm{cl}}(\bm{x}) 
  R^{\dagger} T^{\dagger}\right].
  \label{eq:zero_mode}
\end{align}
The integration over $U$ can be computed by the saddle-point
approximation, which gives the pion mean-field solution. The
integrations over $T$ and $R$ are those over the translational and
rotational zero modes, respectively. They stand respectively for the
translational and rotational transformations that reflect the three
translational zero modes and three rotational ones. $T$ denotes the
dispacement operator that provides the momentum with the nucleon, and
$R$ is the SU(2) unitary matrix. $T$ translates the mean field from
$\bm{x}$ to $\bm{x}-\bm{X}$. The functional 
$\mathcal{F}[U]$ is a generic expression for any correlation function
in the presence of the mean field. 

Having imposed the boundary conditions at infinite Eucliean time
separation, we show that the Fourier transform naturally appears
and the functional integral is reduced to that over the flavor rotation
$R$. Note that the integral over the flavor rotations is 
normalized to unity $\int dR =1$. The integration over $R$ yields the
collective Hamiltonian for a spherical top, which provides $1/N_c$
rotational corrections to the nucleon mass. The eigenfunctions of the
Hamiltonian describe a  baryon state with the quantization condition
$S=T$, so that the lowest baryon state ($S=T=1/2$) is 
the nucleon, and the next state from the rotational excitation
($S=T=3/2$) is identified as the $\Delta$ isobar. The wave functions
$\phi_{B}$ are just the SU(2) Wigner function expressed as 
\begin{align}
\phi_{B}(R) &\equiv \phi^{S=T}_{S_{3}T_{3}}(R) \cr
&= (-1)^{T+T_{3}}\sqrt{2T+1}D^{T=S}_{-T_{3} S_{3}}(R),
\label{eq:sf_wf}
\end{align}
where the baryon quantum numbers are given by 
\begin{align}
B= \{ S=T,S_{3},T_{3} \}.
\label{eq:bary_qn}
\end{align}
Here, $S$ and $T$ denote the baryon spin and isospin quantum numbers,
respectively, and $S_{3}$ and $T_{3}$ represent the corresponding
third components. 

Once the zero-mode quantization is performed, the matrix element of an
effective QCD operator between the baryon states is expressed as  
\begin{align}
\langle B', \bm{p}' | \hat{O} | B, \bm{p} \rangle &=  \int d^{3}\bm{X}
 e^{i  (\bm{p}'-\bm{p}) \cdot \bm{X} } \cr 
&\times \int dR \, \phi^{*}_{B'}(R) \, ... \, \phi_{B}(R).
\label{eq:corre}
\end{align}
The ellipsis in eq.~\eqref{eq:corre} denotes a mean-field
correlation function expressed in terms of the effective QCD
operator~($\hat{O}$) given as functions of the collective
coordinates~($\bm{X}$, $R$). In addition, this correlation function
depends on the configuration of the pion mean
field~($U_{\mathrm{cl}}$) and the one-particle wave functions and
energies. The explicit expression of Eq.~\eqref{eq:corre} will
be presented in Section~\ref{sec:3c}.  

Since we aim at investigating the tensor form factors at large $N_c$,
we want to remark on the normalization of the baryon states and the  
$N_{c}$ scaling of kinematical variables. The masses of the 
lowest-lying baryons are of the order $N_c$, and their mass splitting
is of the order $1/N_c$:   
\begin{align}
&M_{\Delta} = M_{N} \propto N_{c}^1,
  &&M_{\Delta}-M_{N} \propto N^{-1}_{c}. 
\label{eq:Nc_mass}
\end{align}
In the $1/N_{c}$ expansion of the matrix element of the baryon
states~\eqref{eq:corre}, the 3-momentum and energies of the baryon
states $B'$ and $B$ have the following $N_{c}$ scalings: 
\begin{align}
&|\bm{p}'|,|\bm{p}| \propto N_c^{0}, \nonumber \\[1ex]
&p^{\prime 0},p^{0} = M_{N} + \mathcal{O}(N^{-1}_{c}) \propto
 N_c^1. 
\label{eq:3_mo}
\end{align}
In this context, the baryon state in eq.~\eqref{eq:corre} is normalized to
\begin{align}
\langle B', \bm{p}'| B, \bm{p} \rangle&= 2M_N   \delta_{B'B} (2\pi)^{3}
 \delta^{(3)}(\bm{p}'-\bm{p}),  \nonumber \\[1ex]
\delta_{B'B}&= \delta_{S'S} \delta_{T'T} \delta_{S'_3 S_3}\delta_{T_3'T_3}.
\label{eq:bary_nor}
\end{align}

Note that both the center-of-mass coordinates $\bm{X}(t)$ and the
rotation matrix $R(t)$ in Eq.~\eqref{eq:zero_mode} exhibit a weak time
dependence. The gradual displacement and rotation of the soliton
generate kinetic corrections, which are suppressed in the $1/N_c$
expansion. In this work, thus, we concentrate on the rotational and
translational zero modes to the zeroth-order corrections, so-called
strict large $N_{c}$.  

\subsection{Spin-flavor structure \label{sec:3c}}
We are now in a position to apply the local tensor operator,
\begin{align}
\hat{O} = \bar{\psi}_{q} i \sigma^{\mu \nu} \psi_{q},
\label{eq:op}
\end{align}
to Eq.~\eqref{eq:corre}, and derive the spin-flavor structure for
the matrix element of Eq.~\eqref{eq:op} in the pion mean-field
approach. In Ref.~\cite{Kim:2024ibz} an abstract mean-field approach 
was taken and the spin-flavor structure for the matrix element of the 
non-local/local tensor operator was derived. Based on this, the $N_c$
scalings of ``mean-field'' form factors and large-$N_c$ relations were
established. Indeed, we confirm that the spin-flavor structure
obtained in the current approach is consistent with the results of
Ref.~\cite{Kim:2024ibz}. 

In the mean-field picture, the matrix element of the effective local
QCD operator~\eqref{eq:op} is written as 
\begin{align}
&\mathcal{M}^{q}[i \sigma^{\mu \nu}] = \langle B', \bm{p}' |
                \bar{\psi}_{q} \left(0\right) i\sigma^{\mu \nu}
                \psi_{q} \left(0\right) | B, \bm{p} \rangle,  
\label{eq:correlator}
\end{align}
where the baryon state is normalized as given in
Eq.~\eqref{eq:bary_nor} in the leading order of the $1/N_{c}$
expansion. After computing the three-point correlation
function~\eqref{eq:correlator} in the mean-field picture, 
we obtain the matrix elements of the tensor operator~\eqref{eq:op} for
both isoscalar and isovector components in terms of the
first-quantized representation: 
\begin{align}
&\left\{ \begin{array}{c}
\mathcal{M}^{u+d}[i \sigma^{0k}]  \\[2ex] \mathcal{M}^{u-d}[i
           \sigma^{ij}]\end{array} \right\} = 2 M_{N} N_{c}
  \left\{\begin{array}{c}\langle \bm{1} \rangle_{B'B}  \\[2ex] \langle
           D^{3k} \rangle_{B'B} \end{array} \right\} \nonumber \\[2ex] 
&\times  \sum_{n,\mathrm{occ}} \langle n| \left\{\begin{array}{c}
 \gamma^{0}  i \sigma^{0k}  \\[2ex]
 \gamma^{0} i  \sigma^{ij} \end{array}
  \right\} e^{i \bm{\Delta} \cdot \hat{\bm{X}}} | n \rangle
  \bigg{|}_{\mathrm{reg}},  
\label{eq:general_tensor}
\end{align}
The notation $...|_{\mathrm{reg}}$ indicates that a specific
regularization was considered. This procedure depends on the
type of divergence of the matrix element. It will be explored in
Sec.~\ref{sec:3g}.   

The integral over the flavor matrix $R$ in Eq.~\eqref{eq:corre} is
regarded as the matrix element of the spin-flavor operator $O(R)$, 
\begin{align}
& \langle O \rangle_{B'B} \equiv \int dR \, \phi^{*}_{B'}(R) O(R) \phi_{B}(R).
\end{align}
In the leading order of the $1/N_{c}$ expansion, two different types
of the spin-flavor operators, i.e., $\bm{1}$ and $D^{3a}$, appear. The
matrix elements of these operators are obtained to be 
\begin{align}
\langle \bm{1} \rangle_{B^{\prime}B} &= \delta_{B^{\prime} B},
                                       \nonumber \\[1ex] 
\langle D^{3a} \rangle_{B'B} &= -\sqrt{\frac{2S+1}{2S^{\prime}+1}}
                               \nonumber \\[1ex] 
&\times \langle T T_{3}, 1 0 | T^{\prime} T^{\prime}_{3} \rangle
 \langle S S_{3}, 1  a | S^{\prime}  S^{\prime}_{3}
 \rangle . 
\label{eq:sf_me}
\end{align}
Here, $\langle T T_{3}, 1 0 | T^{\prime} T^{\prime}_{3} \rangle$ and
$\langle S S_{3}, 1  a | S^{\prime}  S^{\prime}_{3} \rangle$ represent
the Clebsch-Gordan coefficients. Depending on the choice of the
quantum numbers of the initinal and final states, one derive not only
the $N\to N$ matrix element, but also $N \to \Delta$ and $\Delta \to
\Delta$ matrix elements; see Ref.~\cite{Kim:2023xvw}. 

Equation~\eqref{eq:general_tensor} also includes an infinite tower of
the partial waves $e^{i \bm{\Delta} \cdot \hat{\bm{X}}}$ that comes from
the translational zero mode, where $\hat{\bm{X}}$ represents the 3D
displacement operator. This factor plays an essential role in the
multipole expansion in the mean-field picture. On the other hand,
Eq.~\eqref{eq:general_tensor} is obviously indepdendent of the average
momentum $\bm{P}$, which is natural in the picture of the static mean
field. Using Eq.~\eqref{eq:3_mo}, we find the $N_{c}$ scalings of
these average and difference between inital and final momenta: 
\begin{align}
&P^{0}   \propto N^{1}_{c}, \quad P^{i} \propto
                N^{0}_{c}, \nonumber \\ 
&\Delta^{0} \propto N^{-1}_{c}, \quad \Delta^{i} \propto N^{0}_{c}.  
\label{eq:largeNckin}
\end{align}
These results imply that each partial wave generated by $e^{i
  \bm{\Delta} \cdot \hat{\bm{X}}}$ in Eq.~\eqref{eq:general_tensor}
has the same $N_{c}$ scalings, and that the translational correction
$\bm{P}/2M_N$ is suppressed in $1/N_{c}$ expansion. 

In this context, the matrix element~\eqref{eq:general_tensor} depends
only on $\bm{\Delta}$. Thus, by performing the multipole expansion in
Eq.~\eqref{eq:general_tensor} in powers of $\bm{\Delta}$, we derive
its spin-flavor structure for the isoscalar and isovector components,
respectively 
\begin{subequations}
\label{eq:tensor_FF_largeNc}
\begin{align}
\mathcal{M}^{u+ d}[i\sigma^{0k}] &= Y^{i}_{1} \langle \bm{1}
 \rangle_{B'B} \sqrt{-t}  F^{u+ d}_{\mathrm{mf},1},
\label{eq:tensor_FF_largeNca}
  \\[1ex] 
\mathcal{M}^{u- d}[i\sigma^{ij}] &=-i \epsilon^{ijk} \langle
   D^{3k}\rangle_{B'B} Y_{0} 6M_{N} F^{u- d}_{\mathrm{mf},0} \cr 
&\hspace{-0.7cm}+i \epsilon^{ijk} \langle D^{3m}\rangle_{B'B}
  Y^{km}_{2}  \frac{3t}{2M_{N}} F^{u- d}_{\mathrm{mf},2}
\label{eq:tensor_FF_largeNcb}, 
\end{align}
\end{subequations}
where $F_{\mathrm{mf},0}\equiv F_{\mathrm{mf},0}(t),
F_{\mathrm{mf},1}\equiv F_{\mathrm{mf},1}(t)$, and
$F_{\mathrm{mf},2}\equiv F_{\mathrm{mf},2}(t)$ represent ``mean-field''
form factors. Their subscripts indicate the multipole order $L$ in
$\bm{\Delta}$. We obtain the three independent multipole structures
consistent with the results in Eq.~\eqref{eq:3Dmul}. The explicit
expressions for the soliton form factors are collected in
Sec.~\ref{sec:3e}.   

In principle, $e^{i \bm{\Delta} \cdot \hat{\bm{X}}}$ produces an
infinite number of multipole structures and mean-field form factors. In
the mean-field picture, however, the discrete and hedgehog symmetries
impose constraints on the possible number of multipole structures: i)
The hedgehog symmetry, enforcing the ``minimally generalized''
rotational symmetry, prevents the partial wave from spanning to
infinity, allowing to the order of $L=2$ in the large $N_c$ limit. ii)
Depending on a given effective QCD operator, we find that the parity
allows only even or odd structures in the partial waves. iii) The
time-reversal symmetry restricts the form factors to be real, which 
discards unphysical one in the general
parameterization~\eqref{eq:General_ME_tensor}. These constraints are
consistent with general discrete symmetries imposed on the covariant 
matrix element (cf.~\cite{Schweitzer:2002nm,Schweitzer:2003ms,
  Kim:2024ibz}). Thus we get the same number of multipole structures
in Eq.~\eqref{eq:tensor_FF_largeNc} as in Eq.~\eqref{eq:3Dmul}.  

The baryon matrix element in Eq.~\eqref{eq:tensor_FF_largeNc} can now
be reduced to the nucleon matrix element by taking the quantum number
$S=T=1/2$. From Eq.~\eqref{eq:sf_me} the nucleon matrix elements of
the spin-flavor operators are derived as 
\begin{align}
  \langle \bm{1} \rangle &= \delta_{S'_{3}S_{3}} \delta_{T'_{3}T_{3}},
                           \nonumber \\[1ex]
\langle D^{3i} \rangle &= -\frac{1}{3} (\tau^{3})_{T'_{3}T_{3}}
                         (\sigma^{i})_{S'_{3}S_{3}} , 
\label{eq:sf_nuc}
\end{align}
with Cartesian components ($i=1,2,3$). From now on we will consider
the proton matrix element, i.e., we set $T_3=1/2$. Substituting the
results of Eq.~\eqref{eq:sf_nuc} for the matrix element of the
spin-flavor operator in Eq.~\eqref{eq:tensor_FF_largeNc}, we arrive at
the matrix element of the tensor operator between the proton states:  
\begin{subequations}
\label{eq:tensor_FF_largeNc_proton}
\begin{align}
  \mathcal{M}^{u+d}[i\sigma^{0k}] &= Y^{i}_{1} \sqrt{-t}
              F^{u+ d}_{\mathrm{mf},1},
\label{eq:tensor_FF_largeNc_protona} \\[1ex]
  \mathcal{M}^{u-d}[i\sigma^{ij}] &=i \epsilon^{ijk} \sigma^{k}
    Y_{0} 2M_{N} F^{u- d}_{\mathrm{mf},0} \cr
&\hspace{-0.7cm}-i \epsilon^{ijk} \sigma^{m} Y^{km}_{2}
\frac{t}{2M_{N}} F^{u- d}_{\mathrm{mf},2},
\label{eq:tensor_FF_largeNc_protonb}
\end{align}
\end{subequations}
where we suppress the notations for the spin polarizations $S'_{3}$
and $S_{3}$ in Eq.~\eqref{eq:tensor_FF_largeNc_proton} to be consistent
with Eq.~\eqref{eq:3Dmul}. Owing to the isospin
symmetry~\eqref{eq:sf_nuc}, we also obtain the results for the 
neutron. The isoscalar component of the matrix element for the neutron
is the same as that for the proton. However, the isovector component
of the matrix element for the neutron will have the opposite sign to
that of the proton matrix element. 

By comparing Eq.~\eqref{eq:3Dmul} with
Eq.~\eqref{eq:tensor_FF_largeNc_proton}, we now observe that there is
one-to-one correspondence between the usual tensor form factors
and the mean-field form factors for a selected flavor component in the
large $N_{c}$ limit. The relations between Eqs.~\eqref{eq:3Dmul} and
\eqref{eq:tensor_FF_largeNc_proton} are given by 
\begin{align}
H^{u-d}_{T}+\frac{t}{6M^{2}_{N}} E^{u-d}_{T}&=F^{u-d}_{\mathrm{mf},0},\cr
H^{u+d}_{T}+E^{u+d}_{T}+2\tilde{H}^{u+d}_{T}&=F^{u+d}_{\mathrm{mf},1},\cr
\frac{1}{2}H^{u-d}_{T}+ E^{u-d}_{T}&=F^{u-d}_{\mathrm{mf},2}.
\label{eq:re_1}
\end{align}
On the other hand, the flavor counterparts of
Eq.~\eqref{eq:tensor_FF_largeNc_proton} appear as the rotational
corrections in the mean-field picture, which are suppressed at least
by one power of $N_{c}$. The flavor counterparts of
Eq.~\eqref{eq:tensor_FF_largeNc_proton} are also connected to the
tensor form factors~\eqref{eq:3Dmul}: 
\begin{align}
H^{u+d}_{T}+\frac{t}{6M^{2}_{N}} E^{u+d}_{T}&=F^{u+d}_{\mathrm{mf},0},\cr
H^{u-d}_{T}+E^{u-d}_{T}+2\tilde{H}^{u-d}_{T}&=F^{u-d}_{\mathrm{mf},1},\cr
\frac{1}{2}H^{u+d}_{T}+ E^{u+d}_{T}&=F^{u+d}_{\mathrm{mf},2},
\label{eq:re_2}
\end{align}
which are obviously zero at strictly large $N_{c}$
(cf.~\eqref{eq:tensor_FF_largeNc_proton}), but are non-zero only if 
the rotational corrections are taken into account. While we exhibit
the explicit expressions of Eq.~\eqref{eq:re_1} in Sec.~\ref{sec:3e},
the rotational corrections~\eqref{eq:re_2} are not considered, because
we focus on the $N_c$ behavior of the tensor form factors in the
current work. The corresponding results will be discussed elsewhere. 

The selection rule for determining the leading flavor components in
the $1/N_{c}$ expansion can also be understood in a general way. The
$N_{c}$ scaling of the baryon matrix element in the $1/N_{c}$
expansion is determined by the isospin-spin quantum numbers of the $t$
channels, so-called ``$I=J$ rule''~\cite{Mattis:1988hf, Mattis:1988hg,
  Lebed:2006us}. This rule states that the leading $N_{c}$ structure
comes from the $I=J$ components. For example, for
Eq.~\eqref{eq:3Dmul_a}, its spin quantum number is
$J=0$, so the isoscalar component $I=0$ appears as the leading
contribution, as derived in
Eq.~\eqref{eq:tensor_FF_largeNc_protona}. On the other 
hand, Eq.~\eqref{eq:3Dmul_b} has $J=1$ for both monopole and
quadrupole multipole structures. Thus, its isovector component $I=1$
should be the leading contribution in the $1/N_c$ expansion, with
which Eq.~\eqref{eq:tensor_FF_largeNc_protonb} coincides. Their flavor
counterparts satisfying $I \neq J$ emerge as subleading contributions.

\subsection{Scaling behavior and relations \label{sec:3d}}
The combinations of the tensor form factors on the LHS in
Eqs.~\eqref{eq:re_1} and \eqref{eq:re_2} are not homogeneous in
$N_{c}$, so they are ``incomplete'' relations in the sense of the
$1/N_{c}$ expansion. To have genuine large-$N_{c}$ relations in
Eqs.~\eqref{eq:re_1} and \eqref{eq:re_2}, information on the
separate $N_c$ scaling of the tensor form factors is required, and we 
have to sort out and get rid of the subleading form factors in the
$1/N_{c}$ expansion. The procedure of finding the relations~[see 
Eqs.~\eqref{eq:re_1} and \eqref{eq:re_2}] refined in the $1/N_c$
expansion is as follows: i) First we obtain the $N_{c}$ scaling of the
mean-field form factors. ii) Then we solve the coupled
Eqs.~\eqref{eq:re_1} and \eqref{eq:re_2} to disentangle the separate
usual tensor form factors, and determine the $N_c$ scaling of
them. iii) Having determined the $N_c$ scaling of the tensor form
factors, we ``reverse the logic'' and express the mean-field form factors
in terms of the tensor form factors. In this way, we can derive the
consistent relations -- optimized in the $1/N_{c}$ expansion -- 
between the mean-field form factors and the usual form factors. 

In the leading order of the $1/N_{c}$ expansion, the matrix
element~\eqref{eq:general_tensor} is of the order $\mathcal{O}(N^{2}_{c})$
for both the isoscalar and isovector components. In addition, we find 
that the partial wave expansion does not affect on the $N_{c}$ scaling
of the matrix element; see Sec~\ref{sec:3c}. Thus the matrix element
must produce a uniform $N_{c}$ scaling of the mean-field (multipole)
form factors -- rooted in the same dynamical origin -- in the partial
wave expansions. After performing the partial wave expansion, we do
obtain the uniform ``natural'' scalings of the mean-field form
factors: 
\begin{align}
 &\left\{  F^{u-d}_{\mathrm{mf},0},
   \frac{F^{u+d}_{\mathrm{mf},1}}{M_N},
   \frac{F^{u-d}_{\mathrm{mf},2}}{M_N^{2}} \right\} \sim \ N^{1}_{c}
   \times \mathrm{function}(t). 
\label{eq:Ncscaling_0}
\end{align}
These form factors are independent of the nucleon mass
[cf.~\eqref{eq:Tensor_FFs}], are dimensional quantities, and have the
same $N_{c}$ scaling. Their dimensions are $[\mathrm{mass}^{-L}]$,
where $L$ is the multipole order. To facilitate comparison with the
tensor form factors, we present the $N_{c}$ scaling in terms of the
dimensionless mean-field form factors $F_{\mathrm{mf},L}$: 
\begin{align}
&\{F^{u-d}_{\mathrm{mf},0},F^{u+d}_{\mathrm{mf},1},F^{u-d}_{\mathrm{mf},2} \} \cr
&\sim \ \{N^{1}_{c}, N^{2}_{c}, N^{3}_{c} \} \times \mathrm{function}(t).
\label{eq:Ncscaling_1}
\end{align}
The $N_{c}$ scaling of Eq.~\eqref{eq:Ncscaling_1} is of the order
$\mathcal{O}(N^{L+1}_{c})$. The differences in the $N_c$ scaling arise 
from the multiplication of the nucleon mass. One should keep in mind
that the $N_{c}$ hierarchy between the form factors in
Eq.~\eqref{eq:Ncscaling_1} does not originate from the
dynamics~(rotational or translational corrections), but from
kinematical adjustment. The flavor counterparts of the form factors in
Eq.~\eqref{eq:Ncscaling_1} are generally suppressed by $1/N_c$,
typically one order smaller than the leading components. The $N_c$
scaling for the flavor counterparts is inferred as: 
\begin{align}
&\{F^{u+d}_{\mathrm{mf},0},F^{u-d}_{\mathrm{mf},1},F^{u+d}_{\mathrm{mf},2}
                \} \cr 
&\sim \ \{N^{0}_{c}, N^{1}_{c}, N^{2}_{c} \} \times
                          \mathrm{function}(t). 
\label{eq:Ncscaling_2}
\end{align}
While these form factors are based on a new dynamical origin, compared
to Eq.~\eqref{eq:Ncscaling_1}, the differences in $N_c$ within
Eq.~\eqref{eq:Ncscaling_2} are due to the kinematical adjustment as 
well. 

Using the $N_{c}$ scalings in Eqs.~\eqref{eq:Ncscaling_1} and
\eqref{eq:Ncscaling_2} and taking into account the $N_{c}$ scaling of
the kinematic variables~\eqref{eq:largeNckin}, we rewrite 
Eqs.~\eqref{eq:re_1} and \eqref{eq:re_2} with respect to the separate
tensor form factors and obtain the following equations for the isovector component 
\begin{align}
  F^{u-d}_{\mathrm{mf},0} - \frac{t}{6M^{2}_{N}}F^{u-d}_{\mathrm{mf},2}
  &= H^{u-d}_{T}, \cr
 F^{u-d}_{\mathrm{mf},2}  &= E^{u-d}_{T}, \cr
 -\frac{1}{2}F^{u-d}_{\mathrm{mf},2}  &= \tilde{H}^{u-d}_{T},
 \label{eq:homo_eq1}
\end{align}
and for the isoscalar component
\begin{align}
  F^{u+d}_{\mathrm{mf},0} - \frac{t}{6M^{2}_{N}}F^{u+d}_{\mathrm{mf},2}
  &= H^{u+d}_{T}, \cr
 F^{u+d}_{\mathrm{mf},2}  &= E^{u+d}_{T}, \cr
                            \frac{1}{2}F^{u+d}_{\mathrm{mf},1}-\frac{1}{2}
                            F^{u+d}_{\mathrm{mf},2}
  &= \tilde{H}^{u+d}_{T}.
 \label{eq:homo_eq2}
\end{align}
The left-handed sides~(LHS) of Eqs.~\eqref{eq:homo_eq1} and \eqref{eq:homo_eq2} are now homogeneous in $N_{c}$ scaling. From Eqs.~\eqref{eq:homo_eq1} and \eqref{eq:homo_eq2}, we can read out
the separate $N_{c}$ scaling of the tensor form factors as follows:  
\begin{subequations}
\label{eq:Ncscaling_FFs}
\begin{align}
&\{H^{u-d}_{T}, E^{u-d}_{T}, \tilde{H}^{u-d}_{T}\} \cr
&\sim \ \{N^{1}_{c}, N^{3}_{c}, N^{3}_{c} \} \times \mathrm{function}(t), \\
&\{H^{u+d}_{T}, E^{u+d}_{T}, \tilde{H}^{u+d}_{T}\} \cr
&\sim \ \{N^{0}_{c}, N^{2}_{c}, N^{2}_{c} \} \times \mathrm{function}(t).
\end{align}
\end{subequations}
In addition to that, from the last two equations in
Eq.~\eqref{eq:homo_eq1}, we obtain the non-trivial relation 
(cf.~\eqref{eq:homo_eq1})
\begin{align}
2\tilde{H}^{u-d}_{T} = - E^{u-d}_{T} + \mathcal{O}(N^{1}_{c}).
\label{eq:largeNc_relation_1_tensor}
\end{align}
This relation can also be understood by inserting
Eq.~\eqref{eq:Ncscaling_FFs} into the second line of
Eq.~\eqref{eq:re_2}. On the right-handed side (RHS), the
$F^{u-d}_{\mathrm{mf},1}$ is of the order $\mathcal{O}(N_{c})$. On the
other hand, on the LHS, the 
$E^{u-d}_{T}$ and $\tilde{H}^{u-d}_{T}$ are of the order
$\mathcal{O}(N^{3}_{c})$, and $H^{u-d}_{T}$ is of the order
$\mathcal{O}(N^{1}_{c})$. To have $\mathcal{O}(N^{1}_{c})$ on the LHS,
there must be a cancellation between $E^{u-d}_{T}$ and
$\tilde{H}^{u-d}_{T}$ not only in the order of
$\mathcal{O}(N^{3}_{c})$ but also in the order of 
$\mathcal{O}(N^{2}_{c})$. Thus, the $N_{c}$ order of the subleading
correction to the relation~\eqref{eq:largeNc_relation_1_tensor} must
be two powers smaller than $E^{u-d}_{T}$ and $\tilde{H}^{u-d}_{T}$. 

Using the logical reverse again, and considering
Eqs.~\eqref{eq:homo_eq1} and \eqref{eq:homo_eq2}, we now express the
mean-field form factors in terms of the usual tensor form factors for
the leading flavor components in the $1/N_{c}$ expansion  
\begin{align}
 H^{u-d}_{T} + \frac{t}{6M^{2}_{N}}  E^{u-d}_{T}
  &=F^{u-d}_{\mathrm{mf},0},  \cr 
 E^{u+d}_{T} + 2 \tilde{H}^{u+d}_{T} &=F^{u+d}_{\mathrm{mf},1}, \cr
E^{u-d}_{T} &=F^{u-d}_{\mathrm{mf},2},
\label{eq:relation1}
\end{align}
and for their flavor counterparts
\begin{align}
 H^{u+d}_{T} + \frac{t}{6M^{2}_{N}}  E^{u+d}_{T} & =
F^{u+d}_{\mathrm{mf},0},  \cr 
H^{u-d}_{T} + E^{u-d}_{T} + 2 \tilde{H}^{u-d}_{T}  &=
F^{u-d}_{\mathrm{mf},1}, \cr
E^{u+d}_{T}  &=F^{u+d}_{\mathrm{mf},2}.
\label{eq:relation2}
\end{align}
In the leading order of the $1/N_{c}$ expansion, the combination of
the LHS in Eq.~\eqref{eq:relation1} can be regarded as alternative
definitions of the tensor form factors that have homogeneous $N_{c}$
scaling. For the flavor counterpart~\eqref{eq:relation2}, we also
derive the similar combination. In the perspective of the $1/N_{c}$
expansion, this new basis has a clear physical interpretation and will
be employed in the rest of discussion in this work.  

\subsection{Tensor form factors in the mean-field picture \label{sec:3e}}
In the leading order of the $1/N_{c}$ expansion, the explicit
expressions for the mean-field form factors in
Eq.~\eqref{eq:tensor_FF_largeNc} are collected as follows: 
\begin{subequations}
\label{eq:Tensor_FFs}
\begin{align}
F^{u-d}_{\mathrm{mf},0} &= - \frac{N_{c}}{9} \sum_{n, \mathrm{occ}} \cr
  & \hspace{-0.2cm}  \times \langle n | \gamma^{0}
    (\bm{\Sigma} \cdot \bm{\tau}) Y_{0} j_{0}(|\bm{\hat{X}}|\sqrt{-t})
    | n \rangle \bigg{|}_{\mathrm{reg}}, \\
F^{u+d}_{\mathrm{mf},1} &=  2M_{N} N_{c}  \sum_{n, \mathrm{occ}} \cr
  &\hspace{-0.2cm} \times \langle n | \gamma^{0}\gamma^{5}
    \Sigma^{i} Y^{i}_{1}  \frac{ i j_{1}(|\bm{\hat{X}}|\sqrt{-t})}{
    \sqrt{-t}} | n \rangle \bigg{|}_{\mathrm{reg}}, \\
F^{u-d}_{\mathrm{mf},2}  &=  -2M^{2}_{N}N_{c} \sum_{n, \mathrm{occ}} \cr
  &\hspace{-0.2cm} \times  \langle n |  \gamma^{0}
    \Sigma^{i} \tau^{j} Y_{2}^{ij}  \frac{j_{2}(|\bm{\hat{X}}|
    \sqrt{-t})}{t} | n \rangle \bigg{|}_{\mathrm{reg}}, 
\end{align}
\end{subequations}
where $j_{L}(|\hat{\bm{X}}|\sqrt{-t})$ denotes the 3D spherical Bessel
functions. The irreducible rank-$L$ tensors $Y_{L}$ are given as the
functions of the 3D angles of the displacement operator
$\hat{\bm{X}}$. As already discussed, the quark states are the
eigenstates of the grandspin $\bm{G}$. By coupling the 
first quantized operator~\eqref{eq:Tensor_FFs} in the flavor, the
spin, and the orbital angular momentum spaces, the operator can now be
characterized by the grandspin quantum number. As shown in
Eq.~\eqref{eq:Tensor_FFs}, these first quantized operators between
the quark states $| n \rangle$ are obviously scalar functions and have
no orientation in the ``minimally generalized'' rotation. Thus the
mean-field form factors in \eqref{eq:Tensor_FFs} are only a function of
the squared momentum transfer $t$ (scalar). 

Before discussing more about the mean-field form factors, we need to
specify the relevant domain of $-t$ for them in the large-$N_{c}$
limit. The squared momentum transfer scale $\Delta^{2} = t =
O(N^{0}_{c})$ (cf.~\eqref{eq:largeNckin}), and its temporal component 
$(\Delta^{0})^{2}$ is a subleading correction to $t$. Thus, the lower
bound of $-t$ in the large-$N_{c}$ limit is determined to be  
\begin{align}
   |\bm{\Delta}|^{2} + \mathcal{O}(N^{-1}_{c}) = -t \geq 0.
\end{align}
It means that the squared momentum transfer always remains in the
spacelike domain. On the other hand, $-t$ should not exceed the scale
of the squared nucleon mass $M^{2}_{N}=\mathcal{O}(N^{2}_{c})$ in the
perspective of the large-$N_{c}$ limit. Thus the upper limit of $-t$
is given by   
\begin{align}
 -t \ll M^{2}_{N} \sim 1~\mathrm{GeV}^{2}.
\end{align}
Considering this $N_c$ bahavior of the momentum transfer, we determine 
the reliable domain for $-t$ at large $N_{c}$ to be
$0\leq-t<1~\mathrm{GeV}^{2}$ . Thus, we can unambiguously take the
forward limit $-t \to 0$. In this limit, the spherical Bessel
function in Eq.~\eqref{eq:Tensor_FFs} becomes  
\begin{align}
j_{L}(|\hat{\bm{X}}|\sqrt{-t}) \underset{t\to 0}{=}
  \frac{(|\hat{\bm{X}}|\sqrt{-t})^{L}}{(2L+1)!!}, &&(L=0,1,2\ldots).  
\label{eq:bessel}
\end{align}
Using this asymptotic behavior of the spherical Bessel 
function~\eqref{eq:bessel}, we show that the expressions of the
mean-field form factors are reduced to 
\begin{subequations}
\label{eq:Tensor_moment}
\begin{align}
  F^{u-d}_{\mathrm{mf},0}(0)&=g^{u-d}_{T} = -
| n \rangle \bigg{|}_{\mathrm{reg}}, \\
F^{u+d}_{\mathrm{mf},1}(0)&=\kappa^{u+d}_{T}  \cr
&=  \frac{2M_{N} N_{c}}{3} \sum_{n, \mathrm{occ}} \langle n |
\gamma^{0}\gamma^{5} i  \bm{\Sigma} \cdot \hat{\bm{X}}
| n \rangle \bigg{|}_{\mathrm{reg}}, \\
  F^{u-d}_{\mathrm{mf},2}(0)&=-Q^{u-d}_{T} =
\frac{2M^{2}_{N}N_{c}}{15} \sum_{n, \mathrm{occ}}  \cr
& \hspace{-1.2cm}\times \langle n |  \gamma^{0}
\left[(\bm{\Sigma} \cdot \bm{\hat{X}})
(\bm{\tau} \cdot \bm{\hat{X}}) - \frac{1}{3} (
\bm{\Sigma} \cdot \bm{\tau} )\bm{\hat{X}}^{2}  \right]
| n \rangle \bigg{|}_{\mathrm{reg}}.
\end{align}
\end{subequations}
From Eq.~\eqref{eq:relation1}, we observe that these
results~\eqref{eq:Tensor_moment} are directly related to the tensor
charge $g^{u-d}_{T}$, the anomalous tensor magnetic moment
$\kappa^{u+d}_{T}$, and the tensor quadrupole moment $Q^{u-d}_{T}$,
respectively.  

To facilitate the study of the chiral properties of the tensor form
factors in Eqs.~\eqref{eq:Tensor_FFs} and \eqref{eq:Tensor_moment} we
need to express them in terms of the corresponding densities in
coordinate space. This can be done by projecting
Eqs.~\eqref{eq:Tensor_FFs} and \eqref{eq:Tensor_moment} onto the   
eigenstates of the displacement operator $|\bm{x}\rangle$. Inserting
the completeness 
\begin{align}
\int d^{3} x  \, | \bm{x} \rangle \langle \bm{x} | = 1,
\end{align}
into Eqs.~\eqref{eq:Tensor_FFs} and \eqref{eq:Tensor_moment}, we
obtain the expressions for the mean-field form factors as follows: 
\begin{subequations}
\label{eq:tensor_dis}
  \begin{align}
    F^{u-d}_{\mathrm{mf},0} &=\int^{\infty}_{0} dr \, 4\pi r^{2} \,
                              j_{0}(r \sqrt{-t} ) \rho^{u-d}_{0T}(r), \\
    F^{u+d}_{1,\mathrm{mf}}&= 3 \int^{\infty}_{0} dr \, 4\pi r^{2} \,
                             \frac{ j_{1}(r \sqrt{-t})}{r\sqrt{-t}}
  \rho^{u+d}_{1T}(r), \\
  F^{u-d}_{\mathrm{mf},2} &=15\int^{\infty}_{0} dr \, 4\pi r^{2} \,
\frac{j_{2}(r \sqrt{-t})}{(r\sqrt{-t} )^{2}} \rho^{u-d}_{2T}(r), 
\end{align}
\end{subequations}
where the explicit expressions for the tensor densities are listed in 
Appendix~\ref{app:a}. In the forward limit $t\to 0$, making use of
Eq.~\eqref{eq:bessel}, we obtain the tensor
charge $g^{u-d}_{T}$, the anomalous tensor magnetic moment
$\kappa^{u+d}_{T}$, and the tensor quadrupole moment $Q^{u-d}_{T}$,
respectively, defined in Eq.~\eqref{eq:Tensor_moment}:
\begin{subequations}
\begin{align}
  F^{u-d}_{\mathrm{mf},0}(0) &= \int^{\infty}_{0} dr \,
4\pi r^{2} \, \rho^{u-d}_{T0}(r) = g^{u-d}_{T}, \\
  F^{u+d}_{\mathrm{mf},1}(0) &= \int^{\infty}_{0} dr \,
4\pi r^{2} \, \rho^{u+d}_{T1}(r) = \kappa^{u+d}_{T}, \\
  F^{u-d}_{\mathrm{mf},2}(0) &= \int^{\infty}_{0} dr \,
4\pi r^{2} \, \rho^{u-d}_{T2}(r) = -Q^{u-d}_{T}.
\end{align}
\end{subequations}

\subsection{Discrete level contribution \label{sec:3f}}
The tensor form factors, as discussed in Eqs.~\eqref{eq:energy}
and~\eqref{eq:Tensor_FFs}, comprise the contributions from both the
discrete level~(lev) and negative continuum levels~(Dirac sea)
states. The sum of these two contributions is denoted as ``occ =
discrete level + Dirac sea''. We begin the analysis by considering the 
contribution of the discrete level. The discrete level comes from 
the $G=0$ and $\Pi= +$ state of the quark spectrum. The discrete-level 
wave function takes the following form: 
\begin{align}
  \langle \bm{x} | \mathrm{lev} \rangle = \Phi_{\rm lev} (\bm{x})
  &= \frac{1}{\sqrt{4\pi}}
\left(
\begin{array}{r} f_0 (r) \\[1ex] 
\displaystyle
-i \frac{\bm{x}\bm{\sigma}}{r} \, f_1 (r) 
\end{array} \right) \chi ,
\label{level_spinor}
\end{align}
The discrete-level wave function $\Phi_{\rm lev} (\bm{x})$
incorporates $\chi$, a spinor--isospinor wave function, which 
satisfies the hedgehog condition $(\bm{\sigma} + \bm{\tau}) \chi = 0$,
with the normalization $\chi^\dagger \chi = 1$. The radial components
of the wave function, denoted as $f_0$ and $f_1$, are governed by the 
following equation in the chiral limit $(m = 0)$: 
\begin{align}
  &\left(\begin{array}{cc} M \cos{P(r)} & \displaystyle
-\frac{\partial}{\partial r} - \frac{2}{r} + M \sin{P(r)} \\  
           \displaystyle \frac{\partial}{\partial r} + M\sin{P(r)}
& - M\cos{P(r)} \end{array}\right)
  \cr
&\times \left(\begin{array}{c} f_0(r)  \\[1ex] f_1(r) \end{array}
  \right)= E_{\mathrm{lev}} \left(\begin{array}{c} f_0(r)  \\[1ex]
f_1(r) \end{array}\right),
\label{level_radial}
\end{align}
where $P(r)$ is the profile function of the pion mean field, while
$E_\mathrm{lev}$ represents the level-quark eigenenergy. The
radial wave functions are normalized as 
\begin{align}
\int^{\infty}_{0} dr \, r^{2} \left[f_0^2(r) + f_1^2(r) \right] = 1.
\label{level_normalization}
\end{align}
The subscripts of the quark wave functions $f_0(r)$ and $f_1(r)$
correspond to the orbital angular momentum quantum numbers $L=0$ and
$L=1$, respectively. In the physical mean-field profile, the lower
component $f_1$ accounts for approximately $20\%$ of the normalization
integral~\eqref{level_normalization}, indicating that relativistic
contributions are small but non-negligible. 

Using the discrete-level wave functions~ \eqref{level_spinor}, we
evaluate the first-quantized matrix elements in
Eq.~\eqref{eq:Tensor_FFs}. The resulting discrete-level contributions
to the tensor form factors are then expressed as 
\begin{subequations}
\label{eq:lev}
\begin{align}
  F^{u-d}_{\mathrm{mf},0}[\mathrm{lev}]  &=
\frac{N_{c}}{3} \int^{\infty}_{0} dr j_{0}(r\sqrt{-t}),   \cr
  &\times r^{2} \left[f^{2}_{0}(r) + \frac{1}{3}f^{2}_{1}(r)\right],
    \label{eq:leva}\\
  F^{u+d}_{\mathrm{mf},1}[\mathrm{lev}] &=
-4M_{N}N_{c} \int^{\infty}_{0} dr
\frac{j_{1}(r\sqrt{-t})}{r\sqrt{-t}},
\nonumber \\[1ex]
 &\times r^{3} \left[f_{0}(r)f_{1}(r)\right],
\label{eq:levb}\\
  F^{u-d}_{\mathrm{mf},2}[\mathrm{lev}] &=
\frac{8}{3}M^{2}_{N}N_{c} \int^{\infty}_{0} dr
\frac{j_{2}(r\sqrt{-t})}{(r\sqrt{-t})^{2}}  \nonumber \\[1ex]
&\times r^{4}\left[f^{2}_{1}(r)\right].
\label{eq:levc}
\end{align}
\end{subequations}
Using \eqref{eq:bessel}, we can simplify the discrete-level
contributions to the tensor form factors in the forward limit~($t \to
0$):  
\begin{subequations}
\label{eq:lev_limit}
\begin{align}
  &F^{u-d}_{\mathrm{mf},0}(0)[\mathrm{lev}] =
    \frac{N_{c}}{3} \int^{\infty}_{0} dr  \, r^{2}
    \left[f^{2}_{0}(r) + \frac{1}{3}f^{2}_{1}(r)\right],\\
  &F^{u+d}_{\mathrm{mf},1}(0)[\mathrm{lev}] =
    -\frac{4}{3}M_{N}N_{c} \int^{\infty}_{0} dr \,
    r^{3} \left[f_{0}(r)f_{1}(r)\right],\\
  &F^{u-d}_{\mathrm{mf},2}(0)[\mathrm{lev}] =
    \frac{8}{45}M^{2}_{N}N_{c} \int^{\infty}_{0} dr \,
    r^{4} \left[f^{2}_{1}(r)\right].
\end{align}
\end{subequations}
We see that the monopole mean-field form factor~\eqref{eq:leva}
originates from both the non-relativistic $f_{0}$ and relativistic
$f_{1}$ wave functions. On the other hand, the dipole form
factor~\eqref{eq:levb} arises from the mixing between the $f_{0}$ and
$f_{1}$ wave functions, and the pure $f_{1}$ wave function is
responsible for the quadrupole form factor~\eqref{eq:levc}. This
observation emphasizes that it is crucial to consider the relativistic
effects on the higher multipole structures. 

The above argument can clearly be demonstrated by considering an
extreme limit. In the mean-field picture, the non-relativistic~(NR)
limit can be reached by shrinking the average size of the pion mean
field. Then the quarks interact loosely with the 
mean field in 3D space, so that we paint the non-relativistic 
picture. In this limit, the $|f_{1}|$ over $r$ vanishes as expected,
and the discrete-level energy is reduced to the dynamical quark mass,
i.e. $E_{\mathrm{lev}}\approx M$
\begin{align}
|f_{1}(r)| \underset{\text{NR limit}}{\to} 0, \quad E_{\mathrm{lev}}
  \underset{\text{NR limit}}{\to} M. 
\label{eq:NR_1}
\end{align}
This means that the quarks are very weakly bound, almost like free
quarks. The normalization of the quark wave function in this limit 
becomes 
\begin{align}
\int^{\infty}_{0} dr \, r^{2} f_0^2(r)  = 1.
\label{eq:NR_2}
\end{align}
Under these conditions~\eqref{eq:NR_1} and \eqref{eq:NR_2}, the
mean-field form factors~\eqref{eq:lev} are remarkably simplified as 
\begin{subequations}
\label{eq:NR_dis}
\begin{align}
  F^{u-d}_{\mathrm{mf},0}(0)[\mathrm{lev}]  &=
  \frac{N_{c}}{3} \int^{\infty}_{0}  dr \, r^{2}  f^{2}_{0}(r) \cr
&= \frac{N_{c}}{3} , \label{eq:NR_disa}\\ 
F^{u+d}_{\mathrm{mf},1}(0)[\mathrm{lev}] &= 0, \label{eq:NR_disb}  \\
F^{u-d}_{\mathrm{mf},2}(0)[\mathrm{lev}] &= 0. \label{eq:NR_disc}
\end{align}
\end{subequations}
These results demonstrate that in the NR limit, only the tensor
charge~\eqref{eq:NR_disa} survives with a constant value\footnote{In
  the non-relativistic quark model, the tensor charge is the same as
  the axial charge $g^{u-d}_{A}$, and is known as $(N_{c}+2)/3$. In the
  leading order of the $1/N_{c}$ expansion only part of the results
  $N_{c}/3$ can be explained. The rest of them $\sim2/3$ arises from
  the subleading order contributions (see Refs.~\cite{Kim:1995bq,
    Kim:1996vk}). In this work, we restrict ourselves into the strict
  large $N_{c}$.} $g^{u-d}_{T}[\mathrm{lev}] = N_{c}/3$, while both 
the dipole $\kappa^{u+d}_{T}[\mathrm{lev}] = 0$ and quadrupole
$Q^{u-d}_{T}[\mathrm{lev}] = 0$ moments vanish. The NR limit 
highlights the significance of relativistic effects to describe 
the tensor structure of nucleon, particularly for higher-order
multipoles.  

\subsection{Gradient expansion \label{sec:3g}}
We now examine how the Dirac sea affects the mean-field form
factors. The Dirac sea contribution, unlike the discrete level
contribution, cannot be expressed analytically. It involves the
summation of countless overlapping quark wave functions, each
characterized by quantum numbers (such as energy, grandspin, and
parity). Before we compute the Dirac-sea contributions, we examine the
behavior of the tensor form factors by using the gradient expansion,
which is valid in the limit of the large mean-field size. 
This is called the Skyrme limit, since the expressions in the 
nonvanishing leading-order are identical with the results from the
simplest version of the Skyrme model that contains only the Weinberg 
and Gasser-Leutwyler (Skryme) Lagrangians~\cite{Adkins:1983ya}. Thus,
the current theoretical framework has a merit of interpolating between
the NR and Skyrme limits~\cite{Praszalowicz:1995vi,
  Praszalowicz:1998jm}.    

The gradient expansion can be performed by the following procedure. 
Since a three-point correlation function is given in terms of the
quark propagator given as   
\begin{align}
G(x,y|U) = -\langle x| \frac{1}{i\slashed{\partial}_{\hat{x}} - M
  U^{\gamma_{5}}(\hat{x})} | y \rangle,
\end{align} 
we can expand it in powers of the gradient of the chiral field
($\partial U\ll M$), whose expansion is also known as the 
chiral expansion, since it is compatible with the expansion in powers
of the pion momentum. By considering the nonvanishing leading term in
the expansion, we can evaluate the nucleon matrix element in terms of
the chiral field. The gradient expansion yields analytical expressions
for the nucleon matrix elements, approximating the total matrix
element including both the discrete level and the Dirac sea
contributions. They reveal the chiral properties~(large distance
behavior) of the corresponding nucleon matrix element in the large
$N_{c}$ limit.  

The gradient expansion provides us with two main advantages. First, as
discussed above, it provides theoretical insight into chiral
properties of the 3D densities in the long range  
regime~\eqref{eq:tensor_dis} (so-called the Yukawa tail).
Understanding the large $r$ behavior allows for extrapolation of 3D
distributions, thereby mitigating unwanted finite box effects in
numerical calculation. Second, the low-energy constants for effective
chiral Lagrangian are explicitly derived by performing the 
quark-loop integral, revealing the UV divergence pattern of a relevant
physical quantity. Thus, we can determine whether we have to introduce 
a regularization scheme to tame the corresponding UV
divergence~\cite{Diakonov:1996sr}~\footnote{For example, while the
  pion decay constant is logarithmically divergent, the
  Gasser-Leutwyler low-energy constants for the effective chiral
  Lagrangion to the order $\mathcal{O}(p^4)$ are UV
  stable~\cite{Chan:1986jq, Diakonov:1987ty}. A similar consideration
  can be done also in the case of various nucleonic observables.}.  

Keeping these features in mind, we apply the gradient expansion to the
nucleon matrix element of the tensor operator~\eqref{eq:correlator}, 
and obtain the leading-order contributions to the mean-field form
factors in the chiral limit ($m_{\pi}=0$) 
and in the $1/N_{c}$ and chiral expansions \footnote{
We started with the effective chiral action, as described in
Eq.~\eqref{eq:action}, assuming a zero current quark mass ($\hat{m} =
0$). We then performed a gradient expansion, which finally leads to
the expression given in Eq.~\eqref{eq:grad}. It is important to note
that the inclusion of effects from explicit chiral symmetry breaking in the
effective chiral action may introduce additional contributions to the 
current study~\eqref{eq:grad}. However, for the sake of simplicity in
this study, we have not included these effects.}: 
\begin{align}
  F^{u-d}_{\mathrm{mf},0}[\mathrm{grad}] &=  -\frac{M N_{c}}{288 \pi^{2}}
  \int d^{3}r j_{0}(r\sqrt{-t}) i \epsilon^{ijk}  \cr
  &\times \tr[ L_{i}L_{j} (\tau^{k} U + U^{\dagger} \tau^{k})
    - (i \leftrightarrow j )], \cr
F^{u+d}_{\mathrm{mf},1}[\mathrm{grad}] &=  \frac{M_{N} M N_{c}}{8
  \pi^{2}}    \int d^{3}r \frac{ j_1(r\sqrt{-t})}{r\sqrt{-t}}
 \epsilon^{ijk} r^{k} \cr
& \times \tr[ (U - U^{\dagger}) L_{i} L_{j}], \cr
F^{u-d}_{\mathrm{mf},2}[\mathrm{grad}] &=
\frac{M^{2}_{N} M N_{c}}{16 \pi^{2}} \int d^{3}r
\frac{ j_2(r\sqrt{-t})}{(r^{2}\sqrt{-t})^{2}}
 (r^{l}r^{k} - \frac{1}{3}\delta^{lk}r^{2} ) \cr
  &\times i \epsilon^{ijl}  \tr[ L_{i}L_{j}
    (\tau^{k} U + U^{\dagger} \tau^{k}) - (i \leftrightarrow j )],
\label{eq:grad}
\end{align}
where the trace, $\mathrm{tr}[...]$, runs over flavor space and the
left chiral current $L$ is defined by 
\begin{align}
L_{\mu} := U^{\dagger} \partial_{\mu} U.
\end{align}
Note that these results arise from the non-divergent quark-loop 
integrals. Thus, all tensor form factors are UV finite, eliminating
the need for any regularization scheme. The physical reason for this
lies in the fact that matrix element of the tensor operator solely
originates from the anomalous part of the fermionic 
determinant\footnote{Reference~\cite{Ledwig:2010zq} employed the
  proper-time regularization scheme to estimate the anomalous tensor
  magnetic moment $\kappa^{u+d}_{T}$. However, they introduced the
  regularization functions in an inconsistent way.}; see
Appendix~\ref{app:a} for details. 

We employ the arctangent profile function with the pion Yukawa tail
introduced: 
\begin{align}
P(r) = 2\arctan \left[\frac{R^{2}}{r^{2}}(1+m_{\pi}r) e^{-m_{\pi} r}
  \right], 
\label{eq:arctan_pro}
\end{align}
where $R$ is the average size of the pion mean field, which can be
considered as a variable. The physical value of $R$ is approximately
given as $MR \approx 1$~\cite{Diakonov:1987ty}. Note that the soliton
size $R$ can be reexpressed in terms of the axial charge $g_{A}=1.26$
and the pion decay constant $f_{\pi}=93$~MeV in the chiral limit (see 
Ref.~\cite{Goeke:2007fp}) 
\begin{align}
R = \sqrt{\frac{3g_{A}}{16\pi f_{\pi}^{2}}} \quad [m_{\pi} = 0].
\end{align}
This expression allows one to connect the present results with 
those from chiral perturbation theory~\cite{Goeke:2007fp,
  Alharazin:2020yjv}. Expanding the profile
function~\eqref{eq:arctan_pro} for large $r$ reproduces the correct
pion Yukawa tail:  
\begin{align}
\label{eq:asym}
P(r) &\underset{r \to \infty}{=} \frac{2R^{2}}{r^{2}}  + ..., \
  &[m_\pi = 0], \cr 
P(r) &\underset{r \to \infty}{=} \frac{2R^{2}}{r^{2}} (1+m_{\pi}r)
       e^{-m_{\pi} r} + ..., \ &[m_\pi \neq 0], 
\end{align}
where the ellipses indicate subleading contributions.

Since the chiral field~\eqref{eq:asym} has a correct large $r$
behavior, it is legitimate to examine the large $r$ behavior of the 3D
distributions~\eqref{eq:tensor_dis} with this chiral
field~\eqref{eq:arctan_pro}. By inserting Eq.~\eqref{eq:asym} into 
Eq.~\eqref{eq:grad}, we obtain the asymptotic forms of the 3D
distributions~\eqref{eq:tensor_dis} for a finite pion mass: 
\begin{subequations}
\label{eq:tail_tensor}
  \begin{align}
  \rho_{0T}(r) &\underset{r \to \infty}{=}
  \frac{2M{N_{\mathrm{c}}}R^{4}}{3\pi^{2} r^{{6}}} e^{-2 m_{\pi}r} \nonumber \\[1ex]
  &\times(3+4m_{\pi}r + m_{\pi}^{2}r^{2}) + ...,
   \\
    \rho_{1T}(r) &\underset{r \to \infty}{=}
    \frac{8M M_{N} {N_{\mathrm{c}}} R^{6}}{3\pi^{2} r^{{7}}}
                   e^{-3m_{\pi}r} \nonumber \\[1ex] 
    &\times(1+ 3 m_{\pi}r + 3m_{\pi}^{2}r^{2}+m_{\pi}^{3}r^{3}) + ...,
     \\
    \rho_{2T}(r) &\underset{r \to \infty}{=}
    \frac{8 M M_{N}^{2} {N_{\mathrm{c}}} R^{4}}{{15}\pi^{2} r^{4}} e^{-2m_{\pi} r} \cr
    &\times\left( 1+\frac{5}{3}m_{\pi}r + \frac{2}{3}m_{\pi}^{2}r^{2} \right) + ....
  \end{align}
\end{subequations}
These expressions show the influence of the finite quark mass on the
long-range behavior of the distributions. In the chiral limit $(m_\pi
\to 0)$, these distributions simplify to 
\begin{subequations}
  \begin{align}
  \rho^{u-d}_{0T}(r)&\underset{r \to \infty}{=} 
  \frac{2M{N_{\mathrm{c}}} R^{4}}{\pi^{2} r^{{6}}},   
   \\ 
    \rho^{u+d}_{1T}(r)&\underset{r \to \infty}{=}
    \frac{8M M_{N}{N_{\mathrm{c}}}R^{6}}{\pi^{2} r^{{7}}},
     \\
    \rho^{u-d}_{2T}(r)&\underset{r \to \infty}{=}
    \frac{8 MM_{N}^{2}{N_{\mathrm{c}}} R^{4}}{{15}\pi^{2} r^{4}}.
  \end{align}
\end{subequations}
In particular, the dipole distribution $\rho_{1T}$ falls off, being   
proportional to $1/r^{7}$ in large $r$, while the monopole
$\rho_{0T}$ and quadrupole $\rho_{2T}$ distributions decrease as
$1/r^{6}$ and $1/r^4$, respectively. Note that the quadrupole
tensor distribution diminishes less than the monopole and dipole
ones, so the strong chiral enhancement is 
expected for the quadrupole tensor moment $Q^{u-d}_{T}$ in the chiral
limit compared to the tensor charge and anomalous tensor magnetic
moment.  

By inserting the arctangent profile function~\eqref{eq:arctan_pro}
into Eq.~\eqref{eq:grad}, the mean-field form factors can be expressed
in terms of $R$ in the chiral limit: 
\begin{subequations}
\label{eq:grad_charge}
\begin{align}
F_{\mathrm{mf},0}^{u-d}(0)[\mathrm{grad}] &= \frac{5 {N_{\mathrm{c}}}M
   R}{{12}\sqrt{2}},\\ 
F_{\mathrm{mf},1}^{u+d}(0)[\mathrm{grad}] &= \frac{4 M_{N}
M{N_{\mathrm{c}}}  R^{2}}{{3}\pi},\\ 
F_{\mathrm{mf},2}^{u-d}(0)[\mathrm{grad}] &= \frac{11 M^{2}_{N}
  M{N_{\mathrm{c}}} R^{3}}{{15}\sqrt{2}}. 
\end{align}
\end{subequations}
These expressions reveal that the multipole mean-field form factors
$F_{\mathrm{mf},L}$ with the multipole order $L$ are proportional to
$R^L$. This relationship indicates that higher multipole form factors
exhibit a stronger dependence on $R$. Consequently, it is crucial to
determine $R$ accurately by a self-consistent calculation, in
particular, for the higher multipole form factors. Utilizing the 
physical profile function with $MR\approx 1$ in
Eq.~\eqref{eq:grad_charge}, we derive the following numerical values 
for the tensor moments: 
\begin{align}
&g^{u-d}_{T}[\mathrm{grad}] = 0.88, \cr
&\kappa^{u+d}_{T}[\mathrm{grad}] = 4.25, \cr 
&Q^{u-d}_{T}[\mathrm{grad}] = -17.38.
\end{align}
where we employ the nucleon mass of
$M_{N}=1170$~MeV~\cite{Diakonov:1987ty}, as determined in the chiral  
limit. 

\section{Numerical results}
So far, we have examined the physical features of the tensor form
factors. We now evaluate the tensor formfactors based on the
self-consistent calculation. Before discussing the numerical results,
we briefly show how we determine the parameters. First, 
the dynamical quark mass is taken to be $M=350$~MeV, which is
approximately determined from the QCD instanton vacuum. Here we employ  
the proper-time regularization scheme with a cutoff mass $\Lambda$. By 
reproducing the experimental value of the pion decay constant
$f_{\pi}=93$~MeV  and the pion mass $m_{\pi}=140$~MeV, we
simultaneously fix the current quark mass $m$ and $\Lambda$ (see 
Refs.~\cite{Christov:1995vm, Goeke:2005fs} for details):
\begin{align}
m = 16~\mathrm{MeV}, \quad \Lambda = 643~\mathrm{MeV},
\label{eq:parameter}
\end{align}
where the magnitude of the UV cutoff mass $\Lambda$ is indeed
comparable to the inverse of the average instanton size
$\bar{\rho}^{-1}\approx 600$~MeV. As mentioned in
Eq.~\eqref{eq:energy}, the classical nucleon mass is a logarithmically
divergent quantity and should be
regularized. With~\eqref{eq:parameter} used, we obtain the classical
nucleon mass to be  
\begin{align}
M_{N} = 1254~\mathrm{MeV}.
\end{align}

Evaluating the nucleonic observables, we find that the finite box size 
brings about numerical uncertainty near the boundary of the
box~\eqref{eq:tensor_dis}. To avoide this uncertainty, we replace the
tail part of the densities with those derived from the gradient
expansion~\eqref{eq:grad} at large $r$ (typically around $1-2$
fm). This process effectively restores the chiral properties in the
numerical results. This method allows for a more accurate numerical
study of the chiral properties of tensor form factors in the large
$N_c$ limit of QCD. It has proven particularly valuable in examining
the stability condition and verifying the negativity of the $D$-term
form factor, as demonstrated in previous studies~\cite{Goeke:2007fp,
  Perevalova:2016dln, Polyakov:2018rew}. 

\begin{table}[htp]
\centering
\setlength{\tabcolsep}{5pt}
\renewcommand{\arraystretch}{1.4}
  \begin{tabular}{c|c c c| c }
      \hline
        \hline
      &Level & Dirac Sea & Total & Gradient ($m_\pi=0$) \\
      \hline
      $g^{u-d}_{T}$ &0.88& 0.11& 0.99 & 0.88\\
      $\kappa^{u+d}_{T}$&6.92& 0.69& 7.61 & 4.25\\
      $-Q^{u-d}_{T}$&3.08& {3.94}& {7.02} & 17.38\\
      \hline
      \hline
  \end{tabular}
\caption{Numerical results for the isovector tensor charge
  $g_T^{u-d}$, isoscalar anomalous tensor magnetic moment
  $\kappa_T^{u+d}$, and isovector tensor quadrupole moment with the 
  physical pion mass $m_\pi=140$~MeV employed. The results are also
  compared to those from the gradient expansion in the chiral limit.}
\label{tab:1}
  \end{table}
We find that the dicrete-level contributions to the tensor charge and
anomalous tensor magnetic moment dominate over the Dirac-continuum
contributions, as shown in Table~\ref{tab:1}.
While the discret level yields $g^{u-d}_T[\mathrm{lev}]=0.88$, the
Dirac continuum contributes to it only by around
10\%~\cite{Kim:1995bq, Kim:1996vk,   Schweitzer:2001sr}. Similarly,
the isoscalar anomalous tensor magnetic moment $\kappa^{u+d}_T$ is
primarily governed by the discrete level
($\kappa^{u+d}_T[\mathrm{lev}]=6.92$). Again, the Dirac continuum
gives only $\approx 10\%$ of $\kappa^{u+d}_T$. On the other hand, the
tensor quadrupole moment exhibits significant Dirac-sea contribution
($\gtrapprox 50\%$), which is in line with the case of the electric
quadrupole moment~\cite{Pascalutsa:2006up} and $D$-term form
factor~\cite{Goeke:2007fp, Polyakov:2018rew}. Note that if one
considers only the valence-quark (discrete-level) contributions, results
for $D$-term is underestimated by the experimental
data~\cite{Burkert:2018bqq}. Thus, the Dirac continuum
(sea quarks) must be considered to confront with the
data. These findings imply that relativistic quantum field-theoretical
approaches are essential for accurate description of the nucleon 
structure, particularly for explaining the quadrupole moments
quantitatively.   

The numerical values of the tensor charge and anomalous tensor
magnetic moment have been previously estimated in
Refs.~\cite{Kim:1995bq, Kim:1996vk} and Ref.~\cite{Ledwig:2010zq},
respectively, within the same mean-field picture but with a different
accuracy. The current results reproduce the tensor charges obtained from 
Refs.~\cite{Kim:1995bq} in the strict large $N_c$ limit. 
Concerning the anomalous tensor magnetic moment~\cite{Ledwig:2010zq}, 
we find that while the discrete-level contributions are in agreement
with those from Ref.~\cite{Ledwig:2010zq}, we have the different
results from the Dirac continuum: 
\begin{align} 
\kappa^{u+d}_{T} [\mathrm{sea}] &= 0.05 \quad
                                  (\text{Regularized~\cite{Ledwig:2010zq}}),
                                  \cr 
\kappa^{u+d}_{T} [\mathrm{sea}] &= 0.69 \quad
                                  (\text{Non-regularized}). 
\end{align}
This discrepancy stems from the inconsistent regularization scheme
employed by Ref.~\cite{Ledwig:2010zq}.  

Reference~\cite{Park:2021ypf} estimated the isovector tensor charge of 
the nucleon to be $g^{u-d}_T = 0.97(3)(2)_{\mathrm{sys}}$, using 
$2+1$-flavor Wilson-clover fermions in lattice QCD. On the other hand, 
Alexandrou et al.~\cite{Alexandrou:2021oih} used the twisted mass 
fermions ($N_f = 2+1+1, m_\pi = 260$ MeV) to obtain $g^{u-d}_T =
1.11(2)$. The current results are in good agreement with these lattice
data. However, quark model predictions overestimate
them~\cite{Tezgin:2024tfh, Pasquini:2005dk}. Considering the fact that
the tensor quadrupole moment is identified as one of the first moment
of the chiral-odd GPDs. A recent investigation in lattice
QCD~\cite{Alexandrou:2021bbo} supports the large $N_{c}$
relations~\eqref{eq:largeNc_relation_1_tensor}, which were for the
first time derived from Ref.~\cite{Kim:2024ibz}. 

The multipole tensor moments were extracted also from GPDs computed
within the MIT bag model~\cite{Tezgin:2024tfh}: $g^{u-d}_{T} \approx
1.38$, $\kappa^{u+d}_{T} \approx 6.06$, $Q^{u-d}_{T} \approx
-5$, which are comparable to those presented above. Notably, 
results for the tensor form factors in the forward limit 
$\tilde{H}_{T}(0) \approx -2.6$ and $E_{T}(0) \approx 6.1$ follow
the large $N_c$ relation~\cite{Tezgin:2024tfh}:   
\begin{align}
2\tilde{H}^{u-d}_{T} \approx -E^{u-d}_{T}. \quad (\text{bag 
  model~\cite{Tezgin:2024tfh}}). 
\end{align}
The light-front quark model with the hypercentral
potential~\cite{Pasquini:2005dk} yields $g^{u-d}_{T} \approx 1.21$ and 
$\kappa^{u+d}_{T} \approx 3.15$. While tensor charges are in close
agreement with the current results, the anomalous tensor magnetic
moment is underestimated by about 50\%, compared with the present
prediction. The basis light-front quantization model 
~\cite{Kaur:2023lun} produces a larger value of the tensor charge
($g^{u-d}_{T} \approx 1.52$) but the value of the anomalous tensor
magnetic moment ($\kappa^{u+d}_{T} \approx 5.64$) is closer to our
result. 

The tensor moments are scale-dependent. Thus, it is of great
significance to consider the scale evolution for the tensor moments.
At the one-loop level, the expression for the scale evolution 
was derived in Refs.~\cite{Kodaira:1979ib, Artru:1989zv}: 
\begin{align}
g^{q}_{T} (\mu^{2}) = \left(
  \frac{\alpha_{s}(\mu^{2})}{\alpha_{s}(\mu^{2}_{0})}
  \right)^{\frac{4}{33-2N_{f}}} g^{q} _{T} (\mu^{2}_{0}),
\label{eq:scale}
\end{align}
which shows a weak dependence on the scale. This equation can also be
applied to the anomalous tensor magnetic moment and quadrupole moments,
since they come from the same tensor operator. As the scale increases
towards infinity $\mu \to \infty$, the tensor charge 
slowly approaches zero. While it is relatively stable
across different scales, this scale dependence does introduce some
uncertainty in our calculations. Since the present effective chiral
theory is based on the QCD instanton vacuum, the intrinsic scale is
naturally determined by the average instanton size, is set at $\mu_{0}
\approx \bar{\rho}^{-1} = 600$~MeV~\cite{Kim:1995bq}. 
To maintain consistency with the large $N_c$ approximation and avoid
additional assumptions, we present the results at the intrinsic scale
given above. For more advanced treatments, two-loop and three-loop
evolution expressions are available in literatures~\cite{Barone:2001sp,
  Manashov:2024fcd}.  

\section{Discussion \label{sec:dis}}
\subsection{Non-relativistic quark model vs. the Skyrme model} 
Previously, we showed that the derivative expansion demonstrates
approximately the Skyrmion picture of the present mean-field theory at
its large average size $R$. If we take the opposite limit, i.e., $R\to
0$, we are able to interpolate between the non-relativistic quark model (NRQM) limit and the Skyrme
limit. Both the limits can be regarded as two extreme limits of the
current pion mean-field theory. 

When $R \ll M^{-1}$, massive quarks become weakly bound by the mean
field, with  discrete-level energy approximate to the dynamical quark
mass, $E_{\mathrm{lev}} \approx M$, and vacuum polarization effects
become negligible ($E_{\mathrm{sea}}\approx 0$). It means that the
mean field approaches a non-relativistic quark limit, i.e, $M_{N} \approx
N_{c} M$. In this regime, it exhibits properties consistent
with the NRQM.

Conversely, when $R \gg M^{-1}$, the discrete-level quarks get
strongly bound, so that the discrete-level energy turns negative,
$E_{\mathrm{lev}}<0$. Then, they finally dive into the Dirac  
sea, and the baryon number of the nucleon comes solely from the
Wess-Zumino-Witten term, which is derived by the gradient expansion of
the imaginary part of the effective chiral action
\eqref{eq:action}. Thus, the baryon number is identified as the
topological winding number, and the topological soliton emerges from
the strongly polarized Dirac-sea vacuum. This picture exactly
corresponds to what Witten suggested~\cite{Witten:1979kh}. In this
limit of $MR\gg 1$, the Dirac-continuum energy approaches to the
nucleon mass ($E_{\mathrm{sea}} \approx M_{N}$). 

\begin{figure}[htp]
\includegraphics[scale=0.28]{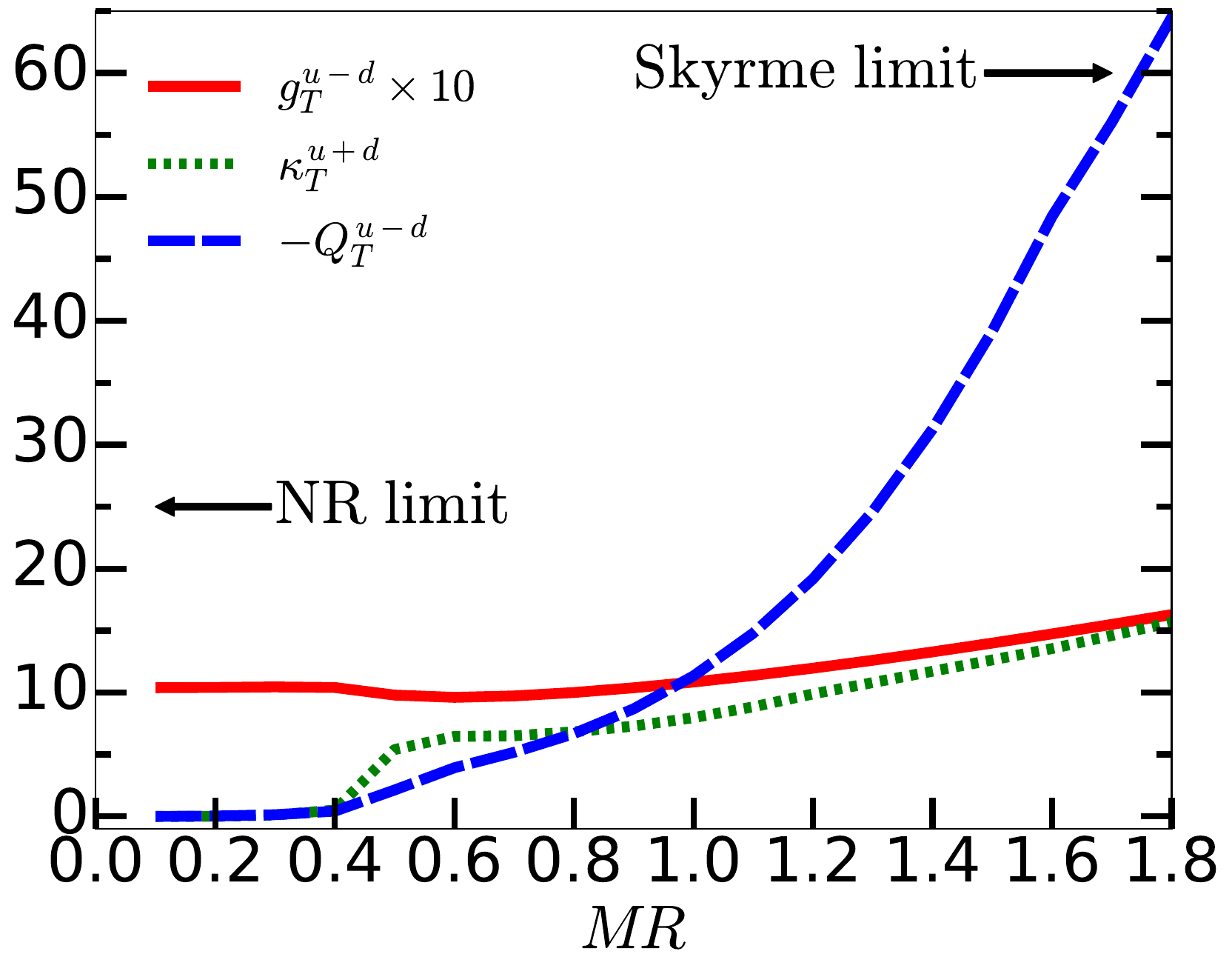}
\caption{Tensor multipole moments computed with the arctangent profile
  function~\eqref{eq:arctan_pro} as a function of the dimensionless
  parameter $MR$. {\it{Solid curve}}: tensor charge. {\it{Dotted
      one}}: anomalous tensor magnetic moment. {\it{Dashed one}}:
  tensor quadrupole moment.}  
\label{fig:1}
\end{figure}
We now compute the tensor mulipole moments, employing the arctangent
profile function instead of the self-consistent one with the
dimensionless parameter $MR$ varied. The results are depicted in
Fig.~\ref{fig:1}, which shows three distinct regimes. For $MR < 0.5$
the quarks are nearly unbound, leading to a plateau. The
self-consistent solution for the pion mean field appears at about
$MR\approx 1$. When $MR$ reaches about 1.5, the discrete level quark
merges with the Dirac sea. Examining these results, we have the 
following observations:  
\begin{enumerate}[label=(\roman*)] 
\item As $R$ decreases below the physical value $(MR <
  1)$, the tensor charge $g^{u-d}_T[\mathrm{lev}]$, proportional to
  $f^2_0 + (1/3)f^2_1$, remains almost stable. This stability
  results from the dominance of the non-relativistic component $f_0$
  in the wave function, with the relativistic component $f_1$
  contributing only a small correction of at most $20\%$ to the tensor
  charge. In addition, the contribution of the Dirac continuum is 
  negligible in this region. Consequently, the value of $g^{u-d}_{T}$
  for $MR \approx 1$ is very close to that from the NRQM~($MR < 0.5$),  
  which suggests the validity of the NRQM prediction for the tensor
  charge $g^{u-d}_{T}$.  

  In contrast, $\kappa^{u+d}_T$ shows a dramatic change around $MR
  \approx 0.5$. This behavior is attributed to its dependence on the
  mixed product of the upper and lower components $(f_0f_1)$ in the
  discrete level contribution, and to the fact that the Dirac
  continuum contributes only marginally to $\kappa_T^{u+d}$. As the
  pion mean field approaches an unbound state $(|f_1| = 0)$,
  $\kappa^{u+d}_T$ decreases sharply due to the decreasing $|f_1|$
  component.  

$Q^{u-d}_T$, which is proportional to $f^2_1$ in the discrete-level 
contribution, exhibits a purely relativistic effect. In addition,
there is a huge Dirac-continuum contribution, which increases rapidly
as $MR$ increases. It reflects the sensitivity of $Q^{u-d}_T$ to
relativistic and Dirac-continnum effects.  

These numerical observations are consistent with the analytical
properties described in Eq.~\eqref{eq:NR_dis}, which predicts that
both $Q^{u-d}_T$ and $\kappa^{u+d}_T$ approach zero in the
non-relativistic limit. 

\item As $R$ increases from the physical one, i.e. $MR>1$, the values
  of the tensor multipole moments increase. The reason can be found in
  the fact that the relativistic effects in the discrete level get 
  stronger and the contribution from the Dirac continuum becomes
  sizable. 

\item In the regime where $MR\gg1$, the behavior of the tensor
  multipole moments is consistent with the predictions from the
  gradient expansion, as explained in
  Eq.~\eqref{eq:grad_charge}. The dependence of these moments on $MR$
  can be characterized as follows: 
\begin{align}
g^{u-d}_{T} &\sim (MR)^{1}, \cr
\kappa^{u+d}_{T} &\sim (M_{N}R)^{1}(MR)^{1}, \cr
Q^{u-d}_{T} &\sim (M_{N}R)^{2}(MR)^{1}.
\end{align}
Here, $M_{N}$ is held constant. These dependence shows distinct
behaviors for different tensor moments as $MR$ increases. The tensor
charge $g^{u-d}_{T}$ exhibits weak dependence on $MR$, whereas 
higher tensor multipole moments are strongly enhanced on account of 
additional $R$-dependence: $k^{u+d}_{T} \propto R^{2}$ and
$Q^{u-d}_{T} \propto R^{3}$. Consequently, these higher moments
are significantly amplified in the large $MR$ region.
\end{enumerate}

This analysis reveals the transition in nucleon structure from
non-relativistic to relativistic regimes as the mean-field size $R$
varies. It indicates how the tensor moments evolve differently with $R$, providing insight into which of the non-relativistic, relativistic, and Dirac-continuum effects is the main source of the values of the tensor moments.

\subsection{Pion mass dependence}
To study the chiral properties of the tensor form factors, we 
compute the tensor moments with the pion mass varied. This 
will shed light on the chiral extrapolation used in lattice
QCD. Moreover, it is necessary to compare the present results with
those of lattice QCD in which an unphysical pion mass is often used.    

To consider different values of the pion mass, we repeat the procedure
of fixing the cutoff and current quark masses, and derive the
corresponding pion mean fields. We strictly follow
Ref.~\cite{Goeke:2005fs} to obtain the pion mass dependence of  
the tensor form factors (see Section~\ref{sec:dis}).  
\begin{table}[htp]
\centering
\setlength{\tabcolsep}{8pt}
\renewcommand{\arraystretch}{1.2}
      \begin{tabular}{c| c c c}
        \hline
        \hline
        $m_{\pi}$ & $g^{u-d}_{T}$[lev]& $g^{u-d}_{T}$[sea]&$g^{u-d}_{T}$[tot] \\
        \hline
        0 &
        {0.885}&{0.117}&{1.002}\\
        10 & 
        {0.885}&{0.117}&{1.002}\\
        50 &
        {0.884}&{0.114}&{0.998}\\
        140 &
        {0.882}&{0.106}&{0.988}\\
        300 &
        {0.882}&{0.091}&{0.973}\\
        600 &
        {0.897}&{0.065}&{0.962}\\
        \hline
        $m_{\pi}$ & $\kappa^{u+d}_{T}$[lev]& $\kappa^{u+d}_{T}$[sea]&$\kappa^{u+d}_{T}$[tot] \\
        \hline
        0 &
        {6.87}&{0.94}&{7.81}\\
        10 &
        {6.87}&{0.94}&{7.81}\\
        50 &
        {6.88}&{0.88}&{7.76}\\
        140 &
        {6.92}&{0.69}&{7.61}\\
        300 &
        {6.86}&{0.51}&{7.37}\\
        600 &
        {6.66}&{0.43}&{7.09}\\
        \hline
          $m_{\pi}$ & $-Q^{u-d}_{T}$[lev]& $-Q^{u-d}_{T}$[sea]&$-Q^{u-d}_{T}$[tot] \\
          \hline
          0 &
          {3.01}&{{9.58}}&{12.59}\\
          10 &
          {3.01}&{8.40}&{11.42}\\
          50 &
          {3.03}&{6.21}&{9.24}\\
          140 &
          {3.08}&{3.94}&{7.02}\\
          300 &
          {3.01}&{2.73}&{5.74}\\
          600 &
          {2.80}&{2.50}&{5.30}\\
          \hline
          \hline
      \end{tabular}
\caption{The tensor multipole moments as functions of the pion
  masses.}   
\label{tab:2}
      \end{table}
The results for the dependence of the tensor multipole moments on
$m_\pi$ are summarized in Table~\ref{tab:2}. We observe that the 
tensor moments $g^{u-d}_{T}$ and $\kappa^{u+d}_{T}$ show only weak
dependence on the pion mass. As $m_\pi$ increases, the discrete-level
contributions to these tensor moments remain almost 
unchanged. However, the Dirac sea contributions diminishes as
$m_\pi$ increases. This can be understood that the heavier pion
suppresses the mean field at large $r$. Thus, the discrete-level
contributions remain dominant for $g^{u-d}_{T}$ and
$\kappa^{u-d}_{T}$, resulting in relatively stable total tensor
moments with the pion mass changed. 
In contrast, $Q^{u-d}_{T}$ arises purely from the relativistic
effects, as discussed in our analysis of $MR$ dependence. For this
quadrupole moment, the Dirac-continuum contribution dominates over the 
discrete level contribution. Consequently, $Q^{u-d}_{T}$ weakens as
$m_\pi$ increases. This implies that $Q^{u-d}_{T}$ becomes
significantly enhanced in the chiral limit. This observation is indeed  
consistent with the result $Q^{u-d}_{T} = -17.38$ from the gradient
expansion in the chiral limit.  

This feature emphasizes the unique nature of $Q^{u-d}_{T}$ among the 
tensor moments and underlines the importance of chiral dynamics. We
anticipate results from lattice QCD in the near future, which will
reveal the nature of the tensor quadrupole moments.

\subsection{$t$ dependence }
The tensor multipole moments are just the values of the corresponding
tensor form factors at $t=0$. We now consider the $t$-dependence of
the form factors. Due to the large $N_{c}$ hierarchy between the
kinamtical variables, we restrict ourselves into the kinematic region
where $0\leq-t<1~\mathrm{GeV}^{2}$ [cf.~Sec.\ref{sec:3e}]. 

\begin{figure}[htp]
\includegraphics[scale=0.28]{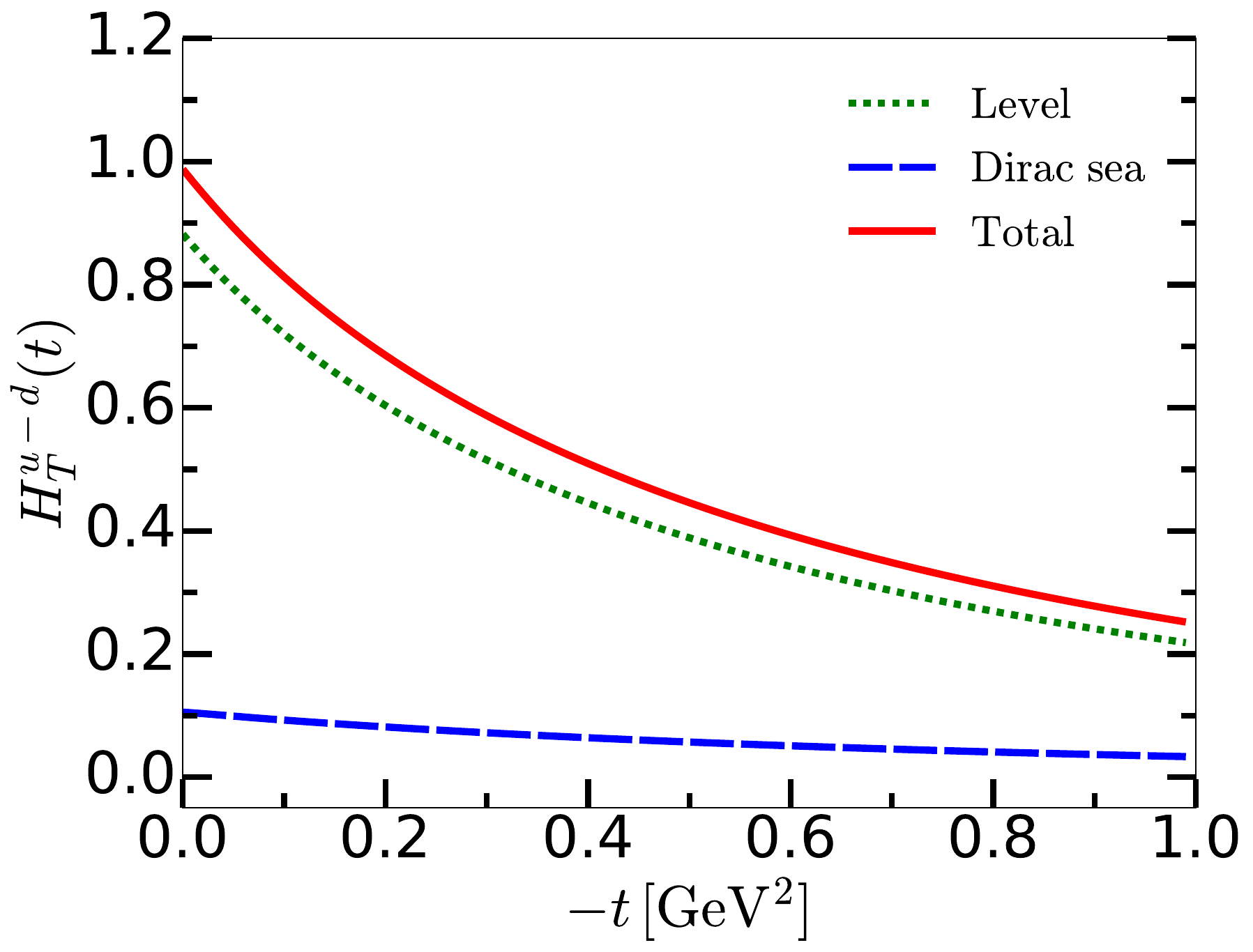}
\includegraphics[scale=0.28]{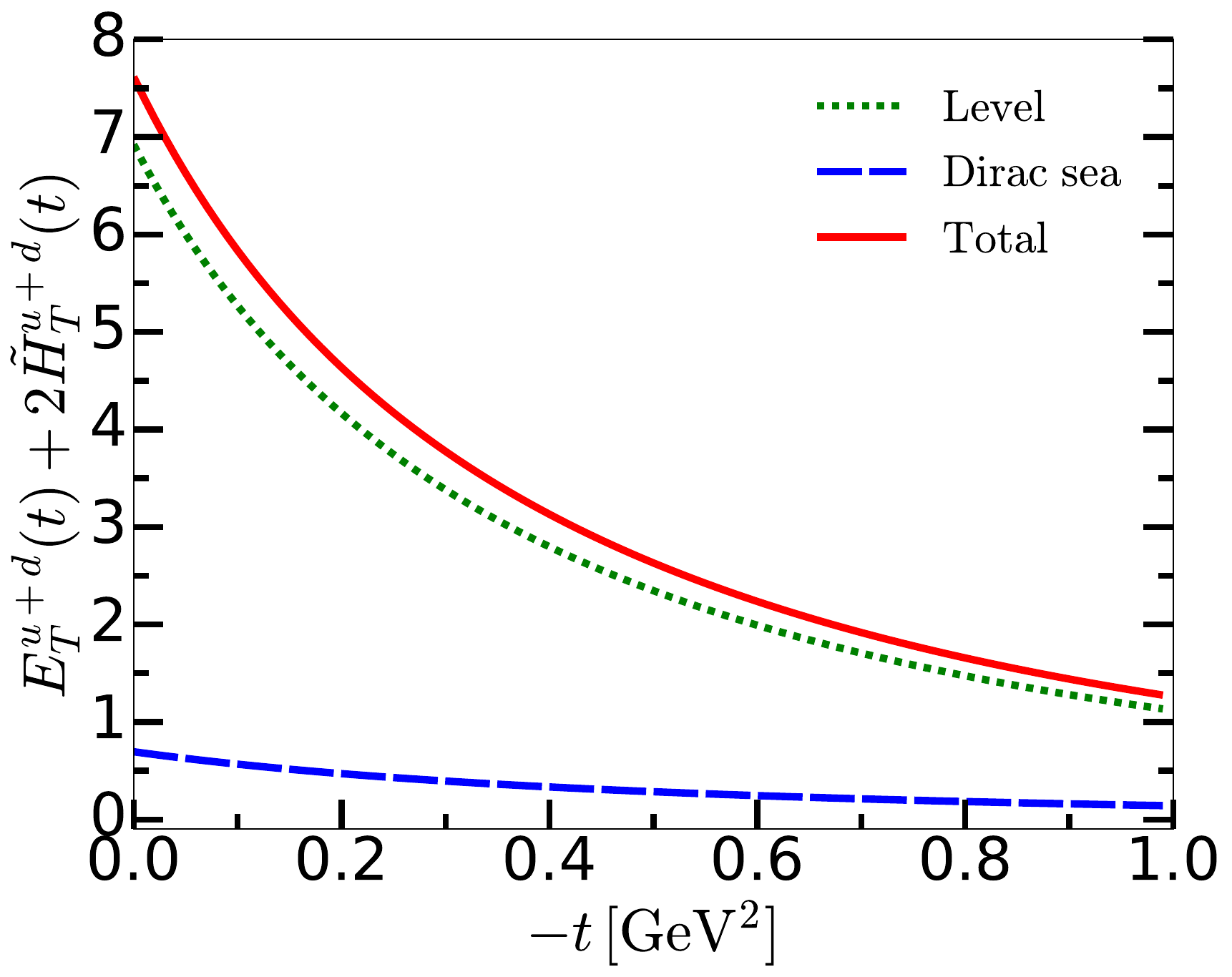}
\includegraphics[scale=0.28]{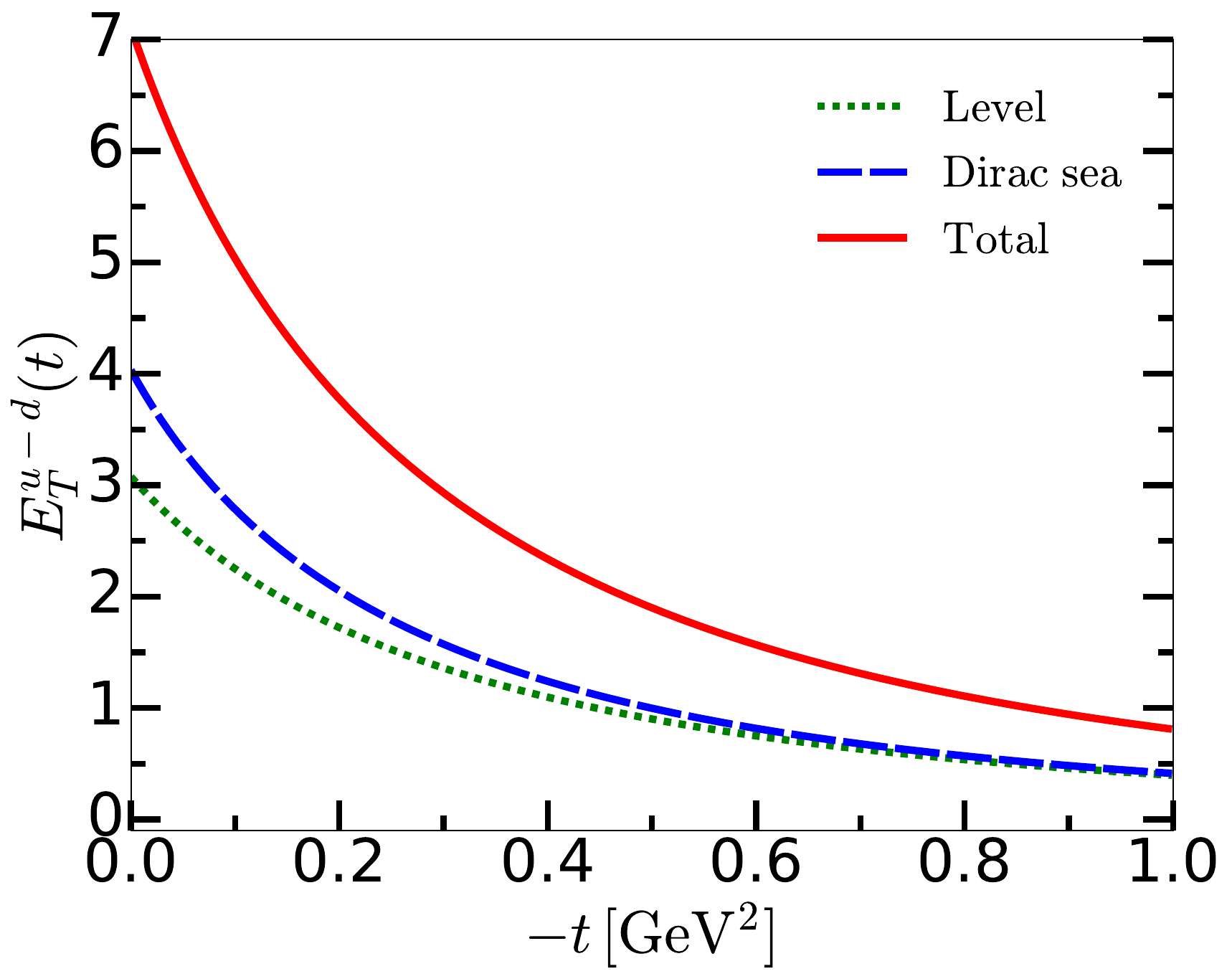}
\caption{The $t$-dependence of the tensor form factors $(H^{u-d}_{T},
  E^{u+d}_{T}+2\tilde{H}^{u+d}_{T}, E^{u-d}_{T})$.
  {\it{Solid curve}}: Total (discrete level + Dirac
  sea) contributions. {\it{Dashed one}}: Dirac sea
  contributions. {\it{Dotted one}}: Discrete level contributions.} 
\label{fig:2}
\end{figure}

Figure~\ref{fig:2} draw the $t$ dependence of the tensor form
factors. For the $H^{u-d}_{T}$ and $E^{u+d}_{T}+2\tilde{H}^{u+d}_{T}$,
the contribution of the Dirac continuum monotonically decreases as $t$
increases. It means that dynamical information is mostly
determined by the discrete-level quarks. However, when it comes to
$E^{u-d}_{T}$, the Dirac continuum takes charge over it. The dashed
curve in the bottom panel of Fig.~\ref{fig:2} shows the contribution
of the Dirac continuum to $E^{u-d}_{T}$, which is larger than the
discrete-level one over the whole $t$ region. 
It reflect the fact that the tensor quadrupole form factor mainly is  
governed by the Dirac continuum. Note that $\tilde{H}^{u-d}_{T}$ is
equal to $-\frac{1}{2}E^{u-d}_{T}$ in the large $N_{c}$ limit. 

\section{Summary and extensions}
In this study, we aimed at investigating the nucleon tensor form
factors in a pion mean-field approach, which is based on the
large $N_{c}$ limit and the effective chiral dynamics resulting from
spontaneous chiral symmetry breaking. At large $N_{c}$ the nucleon
emerges as a mean-field solution of the classical equation of
motion. After the zero-mode quantization, we were able to evaluate the
nucleon matrix elements of the local tensor operator. Our key findings
are as follows:   

\begin{enumerate}[label=(\roman*)]

\item We perform the multipole expansion of the nucleon matrix element
  of the tensor operator in the mean-field picture and derive, in the
  $1/N_{c}$ expansion, its spin-flavor structure and the corresponding
  the mean-field form factors. These form factors are matched
  with the usual tensor form factors by considering the
  comprehensive large-$N_{c}$ analysis. As a result, we determine the
  $N_{c}$ scaling~\eqref{eq:Ncscaling_FFs} of the tensor form factors
  and obtain the non-trivial large $N_{c}$
  relation~\eqref{eq:largeNc_relation_1_tensor}.

\item We evaluated the tensor charge~($g^{u-d}_{T}=0.99$), the
  anomalous tensor magnetic moment~($\kappa^{u+d}_{T}=7.61$), and the 
  tensor quadrupole moment~($Q^{u-d}_{T}=-7.02$) with the physical
  pion mass; see Table~\ref{tab:1}. Our results show that the tensor 
  charge and the anomalous tensor magnetic moment are dominated by
  valence (discrete-level) quark contributions. On the other hand, the
  tensor quadrupole moment comes from the purely relativsitic effects
  and shows  significant sea (Dirac-continuum) quark effects,
  highlighting the importance of chiral dynamics in its description. 

\item Having changed the average size of the pion mean field $R$, we
  investigated how the current   mean-field approach interpolates
  between the non-relativistic quark model and the Skyrme model. We
  focus on two extreme cases: for small $R$ ($R\ll M^{-1}$), the
  present approach yields the picture of the non-relativistic quark
  model, whereas   for large $R$, it is reduced to the Skyrme
  picture. Our results show that the tensor charge $g^{u-d}_{T}$ from
  non-relativistic quark model remains rather accurate due to small
  relativistic and Dirac   sea effects. In the non-relativistic limit,
  the anomalous tensor magnetic moment $\kappa^{u+d}_{T}$ tends to
  zero due to the fact that the discrete-level contribution comes only
  from the relativistic effects and the Dirac-sea effect is
  negligible. However, the tensor   quadrupole moment $Q^{u-d}_{T}$
  not only vanishes in the non-relativistic limit, but also exhibits
  substantial Dirac-sea contributions. This indicates that a
  field-theoretic approach is crucial for an accurate study of the
  quadrupole moment.  

\item We investigates the chiral properties of nucleon tensor form
  factors with the pion mass varied, enabling comparisons with lattice
  QCD results at unphysical pion masses. The results show that tensor
  moments $g^{u-d}_{T}$ and $\kappa^{u+d}_{T}$ exhibit weak pion mass
  dependence, with dominant discrete-level contributions remaining 
  stable. In contrast, the tensor quadrupole moment $Q^{u-d}_{T}$,
  which arises from the relativistic effects, is dominated by Dirac-sea
  contributions, so that it is significantly enhanced near the chiral
  limit. This distinctive behavior of $Q^{u-d}_{T}$ underscores the
  importance of chiral dynamics in tensor moments.  
\end{enumerate}

The spin-flavor symmetry we have explored also allows us to relate the
present results to $N \to \Delta$ transition and $\Delta$ baryon
tensor form factors~\cite{Kim:2023xvw, Kim:2023yhp}. Extending the
present method to flavor SU(3), we can compute the tensor form factors
for the baryon octet and decuplet.   

\section*{Acknowledgments}
JYK wants to express the gratitude to H.D. Son and C. Weiss for
valuable discussion. HChK expresses his gratitude to C{\'e}dric
Lorc{\'e} for valuable discussion and hospitality during his visit
to Le Centre de Physique Th{\'e}orique (CPHT) at {\'E}cole
polytechnique, where part of the present work was done. He is also
grateful to the members of the CPHT for their warm welcome. 
The work was supported by the Basic
Science Research Program through the National Research Foundation of
Korea funded by the Korean government (Ministry of Education, Science
and Technology, MEST), Grant-No. 2021R1A2C2093368 and 2018R1A5A1025563  
(NYGh and HChK). HYW acknowleges France Excellence scholarship through
Campus France funded by the French government (Ministère de l’Europe
et des Aﬀaires Étrangères), Grant No. 141295X. This material is based upon work supported by the U.S. Department of Energy, Office of Science, Office of Nuclear Physics under contract DE-AC05-06OR23177.

\appendix
\section{3D distributions of the tensor structure \label{app:a}}
We list the explicit expressions of the 3D distributions:
\begin{subequations}
\label{eq:3Ddis}
\begin{align}
  \rho_{0T}(r) &= \frac{N_{c}}{9\sqrt{3}}\bigg( \langle v||r \rangle
                 \{O_{0T} \}_{0}\langle r||v\rangle\cr 
  &+\sum_{n}\mathrm{sign}(E_{n})\frac{G(n)}{2}\langle n||r \rangle
    \{O_{0T} \}_{0}\langle r||n\rangle\bigg),\\ 
  \rho_{1T}(r) &=\frac{2i}{\sqrt{3}}M_{N}N_{c}\bigg(
  \langle v||r \rangle \{O_{1T} \}_{0}\langle r||v\rangle\cr
  &-\sum_{n}\mathrm{sign}(E_{n})\frac{G(n)}{2}
  \langle n||r \rangle \{O_{1T} \}_{0}\langle r||n\rangle\bigg),\\
  \rho_{2T}(r)&=\frac{2M_{N}^{2}N_{c}}{15\sqrt{6}}\bigg( \langle v||r \rangle
  \{O_{2T} \}_{0}\langle r||v \rangle\cr
  &-\sum_{n}
  \mathrm{sign}(E_{n})\frac{G(n)}{2} \langle n||r \rangle
  \{O_{2T} \}_{0}\langle r||n \rangle\bigg),
\end{align}
\end{subequations}
where the discrete-level wave function~\eqref{level_spinor} is denoted
by $|v \rangle\equiv | \mathrm{lev} \rangle  $ and $G(n) =
\sqrt{2G_{n}+1}$. The irreducible rank-0 first-quantized operators are
defined by  
\begin{align}
\{O_{0T} \}_{0}&= \{\sigma\otimes\tau \}_{0}\gamma^{0}, \cr
\{O_{1T} \}_{0}&= \{\sqrt{4\pi} r
                 Y_{1}\otimes\sigma\}_{0}\gamma^{5}\gamma^{0}, \cr 
\{O_{2T} \}_{0}&=\left\{ \left\{ \sqrt{4\pi} r^{2}Y_{2}\otimes \sigma
                 \right\}_{1}\otimes\tau \right\}_{0} \gamma^{0}. 
\end{align}
The regularization function coming from the imaginary part of the
fermionic determinant is merely a ``sign'' function in
Eq.~\eqref{eq:3Ddis}, which effectively does nothing to the
regularization. Thus, this result is consistent with the
prediction~\eqref{eq:grad} of the gradient expansion. 

\section{Intermediate cutoff and extrapolation}
The explicit sum over the occupied quark states~\eqref{eq:energy} is
defined by 
\begin{align}
\sum_{n,\mathrm{occ}} \langle n | ... | n \rangle \equiv \sum_{E_n, G,
  G_3, \Pi} \langle n  | ... | n \rangle, 
\end{align}
where the quark state $| n = \{E_{n}, G, G_{3}, \Pi \} \rangle$ is
characterized by the energy, the grand spin, and the parity quantum
number. The numerical calculations involve summation over the quark
energies $ -\infty <E_{n} < \infty$. In this section we will show how
to do the sum over the quark energies numerically. There are two
different way of dealing with this energy sum. 

{\it{Unrefined method~(conventional method):}}
 By introducing a finite range defined by $-k_{\mathrm{max}} < E_{n} <
 k_{\mathrm{max}}$, the energy sum can be implemented numerically. The
 tensor form factors are stabilized when $k_{\mathrm{max}}$ reaches
 approximately $4$~GeV. As explored in the gradient expansion, the
 quark-loop integral is UV finite. All results in this work are
 obtained in this way.

\begin{figure}[htp]
\centering
\includegraphics[scale=0.28]{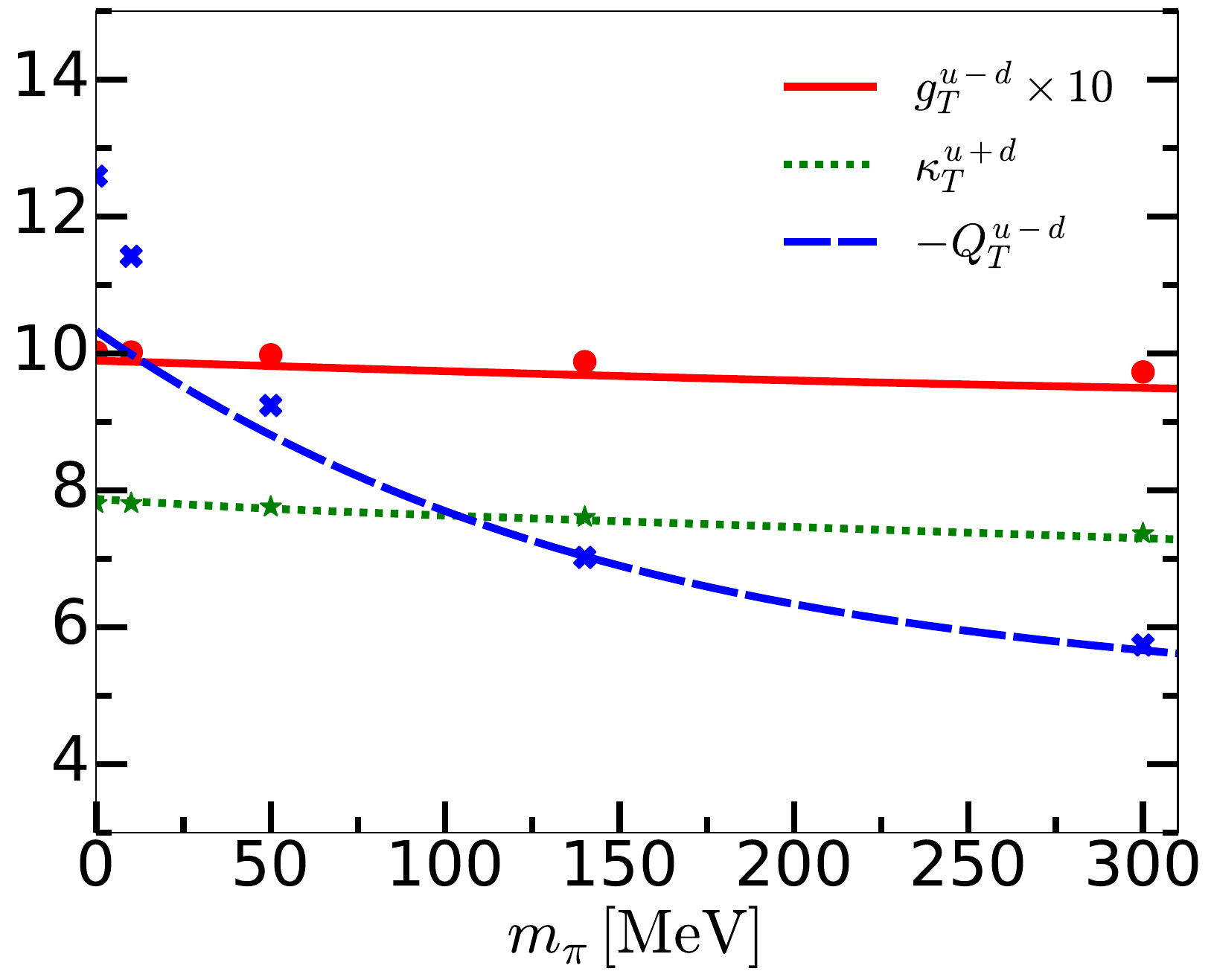}
\caption{Tensor multipole moments as a function of the pion mass: 
  Extrapolated values to $\Lambda_{\mathrm{int}} \to \infty$ and
  $m_{\pi} \to 0$ are compared with the unrefined
  ones. Circles~($\bullet$), stars~($\star$), and crosses~($\times$)
  denote the values of the tensor charge, anomalous tensor magnetic
  moment, and tensor quadrupole moment, respectively (unrefined method). Solid, dotted,
  and dashed curves depict the corresponding extrapolated functions (refined method).} 
\label{fig:3}
\end{figure}

{\it{Refined method:}}
On the other hand, by introducing a varied intermediate cutoff
$\Lambda_{\mathrm{int}}$ instead of $k_{\mathrm{max}}$, one evaluates
the tensor factors for a given range $- \Lambda_{\mathrm{int}} < E_{n}
<\Lambda_{\mathrm{int}}$ where $\Lambda_{\mathrm{int}} \ll
k_{\mathrm{max}}$. Then extrapolate these results for the tensor form
factors to those with $\Lambda_{\mathrm{int}} \to \infty$. This
approach may allow a more rigorous treatment of the UV behavior of the
quark momenta. However, as noted in Ref.~\cite{Schweitzer:2001sr}, the
intermediate cutoff leads to an inequivalence between occupied and
non-occupied sums: 
\begin{align}
\sum_{n,\mathrm{occ}} \langle n | O | n \rangle \neq
  -\sum_{n,\mathrm{non}}  \langle n | O | n \rangle 
\end{align}
This equivalence is necessary for maintaining the correct positivity
condition for parton distribution functions. The inequivalence arises
due to the finite box size and is suppressed when $m_{\pi}
D_{\mathrm{box}} \ll 1$. 
To mitigate these artifacts, the following procedure is employed: 
\begin{enumerate}[label=(\roman*)]
\item Evaluate tensor distributions by varying intermediate cutoffs
  $\Lambda_{\mathrm{int}} \ll k_{\mathrm{max}}$ where equivalence is
  preserved (i.e., fixed large pion mass $m_{\pi}\sim 100 - 200$ MeV) 

\item Extrapolate to $\Lambda_{\mathrm{int}} \to \infty$

\item Repeat for different pion masses and extrapolate to a small pion
  mass. 
\end{enumerate}
In this way we obtain the extrapolated tensor multipole moments at
$\Lambda_{\mathrm{int}} \to \infty$ and $m_{\pi} \to 0$, using the
least squares fit. As shown in Fig.~\ref{fig:3}, the analysis shows
that the extrapolation method yields results for $g^{u-d}_{T}$ and
$\kappa^{u+d}_{T}$ that are consistent with those obtained without
intermediate cutoff refinement. The real problem, however, arises from
the tensor quadrupole moment $Q^{u-d}_{T}$. Like the usual quadrupole
observables (electric quadrupole moment, $D$ term), it depends
strongly on the pion Yukawa tail, which implies that the strong
enhancement is expected at the small pion mass. However, the allowed
pion mass when introducing the intermediate cutoff is about $m_{\pi}
\geq 100$~MeV. This means that this method does not reflect the
correct behavior of the tensor moments at the small pion mass and
introduces huge uncertainties, especially for $Q^{u-d}_{T}$. Indeed, in 
Fig.~\ref{fig:3} we observe that the results with the extrapolation
underestimate the unrefined results, implying a loss of the chiral
dynamics at small $m_\pi$. Therefore, we prefer the unrefined results
in terms of chiral dynamics. However, this procedure remains important 
for the study of parton distribution functions. 

\bibliography{Tensor_FFs}

\end{document}